\documentclass[aip,jcp,reprint,doublecolumn,showpacs,superscriptaddress,nofootinbib]{revtex4-1}

\usepackage{graphicx}
\usepackage{epsfig}
\usepackage{dcolumn}  % Align table columns on decimal point
\usepackage{bm}     
\usepackage{amssymb}     
\usepackage{tabularx}
\usepackage{color} 
\usepackage{setspace}
\usepackage{cleveref}
\usepackage{mathtools}

\begin{document}

%\title{Active and Passive Transport of Cargo in a Corrugated Channel}
\title{Active and Passive Transport of Cargo in a Corrugated Channel: A Lattice Model Study}

\author{Supravat Dey}
\email{supravat.dey@gmail.com}
\author{Kevin Ching}
\author{Moumita Das}
\email{modsps@rit.edu}
\affiliation{School of Physics and Astronomy, Rochester Institute of Technology, Rochester, New York 14623, USA.}

%\pacs{05.40.Jc}%{Brownian motion}
%\pacs{87.16.aj}%{Lattice models}
%\pacs{87.16.dp}%{Transport, including channels, pores, and lateral diffusion}
%\pacs{87.16.Ka}%{Filaments, microtubules, their networks, and supramolecular assemblies}
%\pacs{05.65.+b}{Self-organized systems}
%\pacs{64.60.Cn}{Order-disorder transformations}
%\pacs{64.75.Xc}{Phase separation and segregation in colloidal systems} 

\date{\today}

\begin{abstract}
Inside cells, cargos such as vesicles and organelles are transported by molecular motors to their correct locations via active motion on cytoskeletal tracks and passive, Brownian diffusion. During the transportation of cargos, motor-cargo complexes (MCC) navigate the confining and crowded environment of the cytoskeletal network and other macromolecules. Motivated by this, we study a minimal two-state model of motor-driven cargo transport in confinement and predict transport properties that can be tested in experiments. We assume that the motion of the MCC is directly affected by the entropic barrier due to confinement if it is in the passive, unbound state, but not in the active, bound state where it moves with a constant bound velocity. We construct a lattice model based on a Fokker Planck description of the two-state system, study it using a kinetic Monte Carlo method and compare our numerical results with analytical expressions for a mean field limit. We find that the effect of confinement strongly depends on the bound velocity and the binding kinetics of the MCC. Confinement effectively reduces the effective diffusivity and average velocity, except when it results in an enhanced average binding rate and thereby leads to a larger average velocity than when unconfined.
% we find the average velocity is further reduced.  when the average binding rate is equal to the constant binding rate without confinement.
%Confinement can further modulate the motor's binding kinetics.

\end{abstract}

\maketitle
%%%%%%%%%%%%%%%%%%%%%%%%%%%%%%%%%%%%%%%%%%%%%%%%%%%%%%%%%%%%%%%%

\section{Introduction} 
Intracellular transport of cargos by molecular motors is critical to development, maintenance, and homeostasis in most eukaryotic cells~\cite{AlbertsBook}. There exist several types of motors that use ATP, the energy currency of the cell, to move cargo through the cell using cytoskeletal tracks. The motors kinesin-1 and cytoplasmic dynein transport cargo using microtubules~\cite{Vale2003,Ross2008,Rossgoldman2008,Gunawardena2003,Hirokawa2005,Julicher1997,Appert2015} while myosin-5 and -6 do so via actin filaments \cite{Schuh2013}. Some motors have direction bias; some carry larger, and some smaller cargos. Examples of intracellular cargo include organelles such as mitochondria, and dysfunctional or damaged protein aggregates that occur in disease states~\cite{Vale2003,Ross2008,Holzbaur,Millecamps2013,Zajac2013}. While the transport of the former is essential to proper functioning of the cell, the latter need to be cleared out of the cell to prevent cell damage and disease progression. Understanding the mechanistic principles underlying intracellular cargo transport will provide insights into the proper functioning of cells, and aid in the creation of new drugs or agents to help regain function in disease states.

Microtubules and actin filaments which provide the pathway for the motors to walk during intracellular transport, are semiflexible biopolymers that are found throughout the cell interior~\cite{AlbertsBook,Newby2013,Pilhofer2011}. Over the past two decades, there have been many studies, both experimental and theoretical, on cargo transport by motors on single microtubules in-vitro. The speed and travel distance of molecular motors on surface-immobilized microtubules is very well understood~\cite{Valentine}; examples include the molecular motor kinesin pulling fluid membranes on a microtubule~\cite{Campas2008}, multiple motors transporting a single cargo~\cite{Klumpp2005}, and motors carrying cargos to multiple targets in neurons~\cite{Newby2009}. Within cells, however, microtubules and actin rarely exist as individual filaments. Instead, they are found as networks of filaments and have very interesting mechanical structure-function properties. Despite decades of studies of motors moving and carrying cargo on single cytoskeletal tracks, cargo transport in complex and dynamic architectures in cells is not well understood. In fact, experimental~\cite{Conway2012,Conway2014, Ahmed2014,Fakhri2014,Guo2014,brangwynne2009} and theoretical studies~\cite{Greulich2010,Neri2013,Ciandrini2014,Ando2015,Hofling2013,Goychuk2014,Goychuk2015} have only recently begun to investigate how intracellular transport is affected by the physical properties of the cytoskeletal network and crowded cellular environments. Notable experimental studies in this area include in-vitro experiments that have studied how crowding of motors \cite{Conway2012} and organization of microtubules within bundles \cite{Conway2014} affect the efficiency of cargo transport. In particular, in \cite{Conway2014} Conway {\it et al.} found that the motion of the cargo being transported is inhibited in a bundle of randomly oriented, closely packed microtubules. Several theoretical models with active and passive transport have also investigated the collective transport properties and the spatial organizations of motors and cargos on single microtubules \cite{Ciandrini2014} and inhomogeneous cytoskeletal networks \cite{Greulich2010,Neri2013,Ando2015}. Despite these advances, there remain many open questions.

%\textcolor{blue}{In vitro experiment~\cite{Conway2012}, the authors have investigated how crowding of motors effects the transportation of a cargo. In ~\cite{Conway2014}, Conway {\it et al.} have studied how the organization of microtubules within bundles can affect the cargo transportation and found that the motion of the cargo gets inhibited in a bundle of randomly oriented closely packed microtubules. Several theoretical models with active and passive transportation have been investigating the collective transport properties and the spatial organizations of motors and cargos on a single microtubule \cite{Ciandrini2014} or on inhomogeneous cytoskeletal network \cite{Greulich2010,Neri2013,Ando2015}.} 

%Notable experimental studies in this area include in-vitro experiments that have studied how crowding of motors \cite{Conway2012} and organization of microtubules within bundles \cite{Conway2014} affect the efficiency of cargo transport. In particular, in \cite{Conway2014} Conway {\it et al.} found that the motion of the cargo being transported is inhibited in a bundle of randomly oriented, closely packed microtubules. Several theoretical models with active and passive transport have also investigated the collective transport properties and the spatial organizations of motors and cargos on single microtubules \cite{Ciandrini2014} and in-homogeneous cytoskeletal networks \cite{Greulich2010,Neri2013,Ando2015}. Despite these advances, there remain many open questions."

Here we ask: How does confinement due to the cytoskeletal network affect motor-driven cargo transport?  We address this question by developing a minimal two-state model that describes cargo transport in the presence of confinement.  The two states are: (i) an active state when the motor-cargo complex (MCC) is attached to the microtubule and moves with a constant speed, and (ii) a passive state when it is unattached and undergoes diffusive motion.  Such two-state models have been useful in elucidating active transport of Brownian particles in confined geometries~\cite{Malgaretti2012, Malgaretti2013}, specifically how the cooperative rectification between geometric constraints and Brownian ratchets impacts net particle motion. The interplay between passive and active transport and confinement, as is common in intracellular transport, however, remains poorly understood. In this paper, we combine a Fokker Planck description with a lattice model framework to study how confinement, and motor dynamics and binding kinetics interact to modify directed transport of cargos by motors. 

The paper is organized as follows. We write down the Fokker Planck Equations (FPE) for the two-state model and propose a lattice model that can capture the physics described by the FPE and reduces to the FPE in the continuum limit. We simulate the lattice model using a kinetic Monte Carlo method, and show that it reproduces known analytical results for passive (diffusive) transport in confinement. Thereafter we investigate the full two state problem in confinement, and calculate transport properties such as mean squared displacement (MSD), average velocity, and effective diffusivity for the MCC, and discuss the implications of our results.  

We want to note that, in this paper, we use the term ``active'' to refer to the driven motion of cargo fueled by ATP-hydrolysis of kinesin motors via a ``Brownian ratchet'' mechanism \cite{Oster}, and implemented as constant velocity motion of an MCC on a microtubule (see Sec.~\ref{sec:model}). It should not be confused with self-propelled motion in soft matter literature. The confinement effect on the motion of a self-peopelled particle studied in \cite{Ghosh2013}.

\section{Model and Method}
\label{sec:model}

\begin{figure}[!t]
\includegraphics[width=0.4\textwidth]{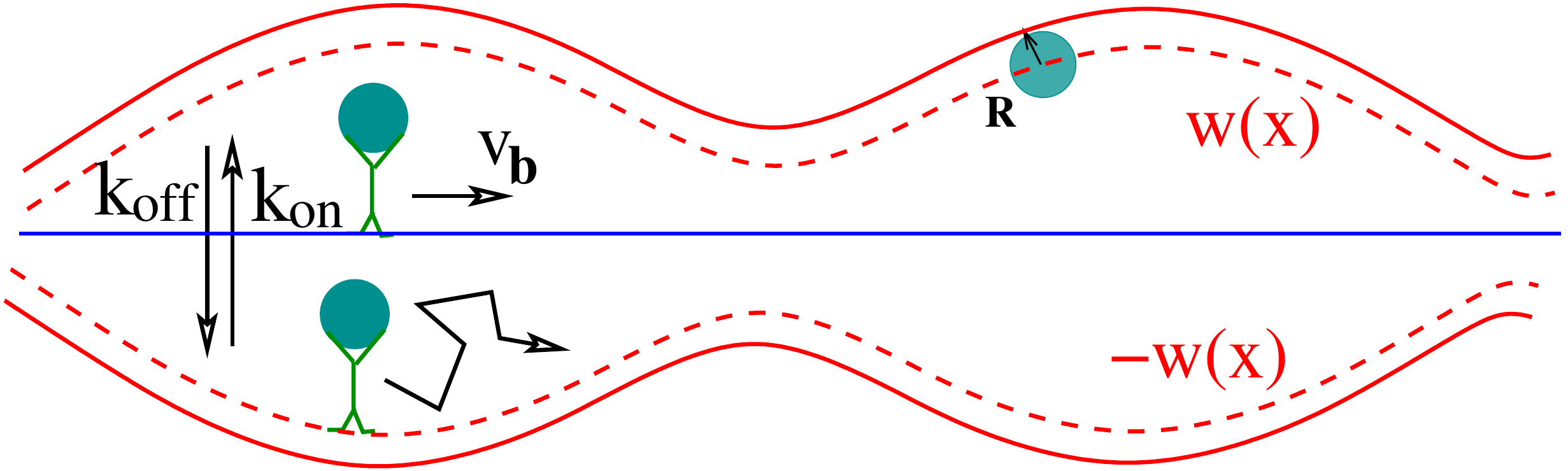}
\includegraphics[width=0.4\textwidth]{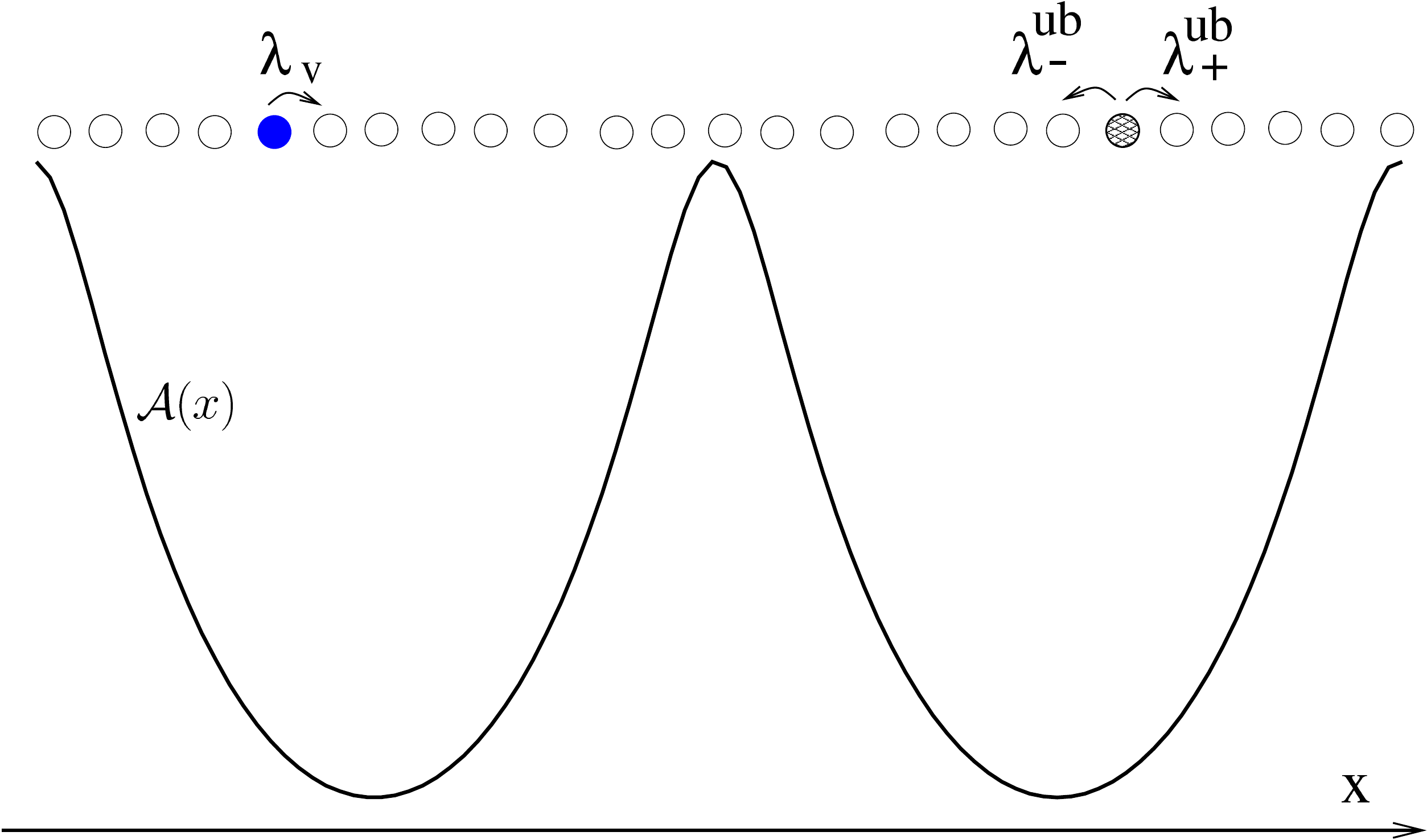}
\caption{Schematic diagram of the model. Top: An MCC (filled circle on stick figure in green) can either bind to a microtubule (straight blue line) with a rate $k_{\rm on}$ and walk with a velocity $v_b$ in the $+x$ direction or it can get detached from the microtubule with a rate $k_{\rm off}$ and diffuse in the corrugated channel. The transverse position of the channel is given by $\pm (R+w(x))$ (solid curved lines in red), for finite size MCC of radius R. For an equivalent point MCC, the space in the transverse direction that is effectively available for diffusion is $\pm w(x)$ (dashed curved lines in red). Bottom: In the lattice model, an MCC either can be in the bound state (filled blue circle) or unbound state (meshed gray circle). A bound MCC only hops in the $+x$ direction, while an unbound MCC hops in both $+$ and $-x$ directions with appropriate rates that depend on the entropy barrier due to the confining wall $w(x)$.} 
\label{fig:model}
\end{figure}

We model and study the active and passive transport of cargo by a motor moving unidirectionally on a microtubule track and confined in a corrugated channel as schematically shown in Fig.~\ref{fig:model}. 
The motor can move micrometer-long distances along the microtubule before detaching. Kinesin motors, a well-characterized family of motor proteins that  move organelles (e.g. mitochondria) and macromolecules (e.g. RNA) in many cell types~\cite{Suetsugu2010} are good examples of such motors. While the confinement faced by an MCC in a live cell is heterogeneous and dynamic, for simplicity, we consider an effective confining channel described by $w(x) = a \sin(2\pi x/L) + b$, where $L$ is the periodicity, $a$ and $b$ control the effective width of the channel, and the effective bottleneck width is given by $2(b-a)$~\cite{Reguera2006}. The effective channel width and periodicity are set by the length scales associated with localized cages and network mesh sizes. 
%For a finite size cargo of radius $R$, the available space in the transverse direction become small, and the effective channel for the cargo can be given by, $w^{\rm eff}(x) = a \sin(2\pi x/L) + (b-R)$. In this paper, when we use the word confining wall we mean the effective confining wall for the cargo, not the actual wall and for conciseness we use same notation $w(x)$ for the effective confining wall without loss of generality. 

The main ingredients of the model are as follows:
\begin{enumerate}
\item Two State Transport: Over long time scales, the MCC alternates between two states: (i) An active state which phenomenologically represents the MCC being bound to a microtubule and moving along the microtubule in the forward direction with a speed $v_b$, and (ii) an passive state where the MCC is detached from the microtubule and undergoes overdamped Brownian motion in the viscus medium of the cytoplasm with a free diffusion constant $D_0$.
 
\item Confinement: The MCC only encounters physical confinement due to the channel walls while undergoing diffusive motion in a viscous medium, i.e. in the passive state. Our model, therefore, incorporates an explicit, microscopic description of the physical confinement due to cytoskeletal networks. This is in contrast to a more coarse-grained, continuum description where interior of the cell is treated as a viscoelastic medium \cite{Goychuk2014,Goychuk2014a,Goychuk2015,Bouzat2014}. The channel is symmetric, and hence we do not expect confinement-induced symmetry-breaking for purely diffusive motion as in some Brownian Ratchet models~\cite{Reimann2002,Hanggi2009}.   

\item Binding Kinetics: The binding rate, $k_{\rm on}(x)$, of the motor can be constant or it can vary inversely with the width of the confining channel in Fig.~\ref{fig:model}. The latter represents the case where tighter confinement leads to greater likelihood of the motor attaching to the microtubule due to increased proximity to binding sites. The unbinding rate, $k_{\rm off}(x)$, of the motor is assumed to be constant. 

\end{enumerate}

We note that the actual width of the confining channel is larger than the size of the MCC. In our formulation, $w(x)$ is the effective channel width (depicted by the dashed curved lines in Fig.~\ref{fig:model}), and the MCC can be thought of as a point particle, with the actual size of the MCC incorporated through the free diffusivity. With this description, our model is valid for a wide range of cargo sizes as long as the bound state is unaffected by the confining channel.

The model predicts three distinct regimes, a diffusive regime at early times, an intermediate sub-diffusive regime, and a ballistic regime at large times, as discussed in detail in the Results section. The timescale for the crossover from diffusive to sub-diffusive motion is set by the diffusion time of the MCC in the corrugated channel before it starts to experience the impact of confinement, while the timescale for the crossover to the ballistic regime is set by the interplay of the binding kinetics and the driven motion of the MCC when bound to the microtubule.

\subsection{Fokker Planck Description}

In our model, the motion of the MCC is directly affected by the confining wall of the channel when it is undergoing Brownian motion in the unbound (passive) state. 
First, let us discuss the motion of an overdamped Brownian particle in a 2D confining channel, with the channel axis along the $x$ direction.
The 2D motion inside the channel can be described by a 1D Fokker-Planck equation, known as the Fick-Jacobs equation~\cite{Zwanzig1992,Reguera2001,Reguera2006},
\begin{eqnarray}
\frac{\partial P(x,t)}{\partial t} &=& D_0 \frac{\partial }{\partial x}\left( e^{-\beta \mathcal{A}(x)} \frac{\partial}{\partial x} e^{\beta \mathcal{A}(x)}P(x,t) \right).
\label{eqn:fpeold}
\end{eqnarray}
Here, $P(x,t)$ represents the probability density at a given position $x$ along the direction of the channel at time $t$,  $D_0$ is the diffusion coefficient in the absence of confinement, and $\beta=1/{k_BT}$, where $k_B$ is the Boltzmann constant and $T$ is the temperature. To derive the above equation, rapid equilibration is assumed in the transverse direction of the channel. This implies that the time scale for longitudinal (axial) motion is very large compared to the equilibration time scale in the transverse direction. Under this assumption, one can successfully integrate out the transverse variable and recast the two-dimensional motion into the above Fick-Jacobs equation \cite{Reguera2001,Reguera2006}. The confinement is incorporated though an effective free energy $\mathcal{A}(x)=\mathcal{V}(x) - T\,S(x)$, where $S(x)$ is the entropy barrier due to confinement and $\mathcal{V}(x)$ is an external energy barrier. The entropy due to the confining wall $w(x)$ is $S(x) =k_B \log(2 w(x)/w_{ave})$, where $w_{ave}=2\int_0^L w(x)\,dx$ is the average width of the channel. In the absence of any external potential, the free energy is purely entropic, $\mathcal{A}(x)=-k_B T\,\log(2 w(x)/w_{ave})$.

Now, we return to the problem of two-state transport. The MCC walks with a velocity $v_b$ when it is bound, diffuses with a free diffusion constant $D_0$ when unbound, and alternates between the two states with rates $k_{\rm off}(x)$ and $k_{\rm on}(x)$, respectively. The Fokker-Planck equation for the probability densities for the bound state $P_{\rm b} (x,t)$ and unbound state $P_{\rm ub}(x,t)$ are given by:
\begin{subequations}
\label{eqn:fpe}
\begin{eqnarray}
\label{eqn:fpeactive}
\!\!\!\frac{\partial P_{\rm b}(x,t)}{\partial t} &=& k_{\rm on}(x) P_{\rm ub}(x,t) - k_{\rm off}(x) P_{\rm b}(x,t) \nonumber \\
&-& v_b \frac{\partial P_{\rm b}(x,t)}{\partial x}, \\
\!\!\!\frac{\partial P_{\rm ub}(x,t)}{\partial t} &=& -k_{\rm on}(x) P_{\rm ub}(x,t) + k_{\rm off}(x) P_{\rm b}(x,t) \nonumber \\
&+&  D_0 \frac{\partial }{\partial x}\left(\!e^{-\beta \mathcal{A}(x)} \frac{\partial}{\partial x} e^{\beta \mathcal{A}(x)}P_{\rm ub}(x,t)\!\right)\!.  
 \label{eqn:fpepassive}
\end{eqnarray}
\end{subequations}

The first two terms in Eqs.~\ref{eqn:fpeactive} and~\ref{eqn:fpepassive} correspond to binding and unbinding transitions respectively. The third term in Eq.~\ref{eqn:fpeactive} represents active motion of the MCC, while the third term in Eq.~\ref{eqn:fpepassive} describes passive motion of the MCC under confinement. Given kinesin is a highly processive motor and can take over a hundred steps along a microtubule before dissociating~\cite{block1990,borisy}, we neglect diffusion in the active state.  However, one can easily incorporate diffusive behavior for other motor types by adding a diffusion term in Eq.~\ref{eqn:fpeactive}. 
The analytical solutions of Eq.~\ref{eqn:fpe} are difficult, and have closed form expressions only in the passive limit~\cite{Reguera2001,Reguera2006} and in a mean field limit for two-state transport discussed later in the paper. We, therefore, construct the corresponding lattice model which reduces to Eq.~\ref{eqn:fpe} in the continuum limit and evolve the system using a kinetic Monte Carlo method as discussed below. 

\subsection{Lattice Model}

We study the dynamics of a two-state MCC described by the continuum Fokker-Plank equation (Eq.~\ref{eqn:fpe}) using an equivalent lattice model. The model is schematically shown in Fig.~\ref{fig:model}, and consists of an MCC on a one-dimensional lattice. The MCC can switch between a bound and an unbound state. The bound MCC can further hop to its forward neighboring site while the unbound MCC can hop to both its backward and forward neighboring sites. The spacing between neighboring lattice sites is $\ell$.  

Consider that the MCC is at the lattice site at position $x$ at time $t$ in a particular state. The transition rates from the unbound state to bound state is $k_{\rm on}(x)$ and from the bound state to unbound state is $k_{\rm off}(x)$. The MCC in the bound state can either hop to its forward neighbor $(x+\ell)$ with rate $\lambda_v(x)$, or it can switch to the unbound state with rate $k_{\rm off}(x)$. The MCC in the unbound state can hop either to its forward neighbor $(x+\ell)$ with rate  $\lambda^{\rm ub}_{+}(x)$, backward neighbor $(x-\ell)$ with rate  $\lambda^{\rm ub}_{-}(x)$, or switch to the bound state with rate $k_{\rm on}(x)$. The master equations describing the time evolution of the probability densities for the bound state $P_{\rm b}(x,t)$ and the unbound state $P_{\rm ub}(x,t)$ for this process are
\begin{widetext}
\begin{subequations}
\label{eqn:mastereqn}
\begin{eqnarray}
\label{eqn:mastereqn1}
\frac{\partial P_{\rm b}(x,t)}{\partial t} &=& k_{\rm on}(x) P_{\rm ub} (x, t) - k_{\rm off}(x) P_{\rm b} (x, t) + \lambda_v(x-\ell) P_{\rm b} (x - \ell, t) - \lambda_v(x) P_{\rm b} (x, t),\\\nonumber
\frac{\partial P_{\rm ub}(x,t)}{\partial t} &=& -k_{\rm on}(x) P_{\rm ub} (x, t) + k_{\rm off}(x) P_{\rm b} (x, t) + \lambda^{\rm ub}_{+} (x - \ell) P_{\rm ub} (x - \ell, t) 
+ \lambda^{\rm ub}_{-}(x + \ell) P_{\rm ub} (x + \ell, t) \\ \nonumber  
&-&  \left(  \lambda^{\rm ub}_{+}(x) + \lambda^{\rm ub}_{-}(x)\right) P_{\rm ub} (x, t).\\
 \label{eqn:mastereqn2}
\end{eqnarray}
\end{subequations}
\end{widetext}

As the bound velocity $v_b$ in our model is independent of position, the bound state hopping rate $\lambda_v(x)$ is also position independent and is given by $\lambda_v(x) = v_b/\ell$. We incorporate the effect of confinement using position dependent hopping rates $\lambda^{\rm ub}_{\pm}(x)$ for the unbound state. The hopping rates $\lambda^{\rm ub}_{\pm}(x)$ depend on the free-energy $\mathcal{A}(x)$ which has a contribution from the entropic barrier due to confinement. In the presence of an external potential, it also has an energy contribution. The hopping rates are given by $\lambda^{\rm ub}_{\pm}(x) = (D_0/\ell^2) \,{\rm e}^{-\beta(\mathcal{A}(x\pm \ell) -\mathcal{A}(x))/2}$. The factor $1/2$ in the exponent ensures local detailed balance condition.
With these choice of rates, for $\ell \rightarrow 0$, the Eq.~\ref{eqn:mastereqn} reduces to Eq.~\ref{eqn:fpe} (see Appendix.~\ref{app:a} for details).

We use a kinetic Monte Carlo algorithm to evolve the system. For the MCC at the lattice site $x$ at time $t$, we choose an event out of all possible events at random with a probability proportional to its rate, and increase the time by $\delta t = 1/\Gamma(x)$, where $\Gamma(x)$ is the total rate. For the bound MCC, the event space consists of a forward hopping event with probability $\lambda_v/\Gamma$, and a transition to the unbound state with probability $k_{\rm off}(x)/\Gamma$, where the total rate in the bound state $\Gamma=k_{\rm off}(x) + \lambda_{v}(x)$.  For the unbound MCC, the event space consists of a hopping event in the forward direction, a hopping event in the backward direction, and a transition to the bound state, with probabilities $\lambda^{\rm ub}_{+}(x)/\Gamma$, $\lambda^{\rm ub}_{-}(x)/\Gamma$, and $k_{\rm on}(x)/\Gamma$ respectively, where the total rate in the bound state is $\Gamma=k_{\rm on}(x) + \lambda^{\rm ub}_{+}(x) + \lambda^{\rm ub}_{-}(x)$. For simplicity and efficiency, the mean of the exponential distribution $\Gamma (x) \exp(-\delta t \,\Gamma (x))$ is used as the time step in our simulations.  Although a time step drawn at random from the exponential distribution would have been more appropriate, we have checked that the choice of the mean does not change any of our results, while it makes the simulation more efficient. This was also verified by one of the authors in a study of a lattice model for ballistic aggregation in~\cite{Dey2011}.
 
It has been shown that the introduction of a position dependent diffusivity, $D(x)=D_0/(1+w^{\prime}(x)^2)^{\alpha}$ (with $\alpha=1/3$ for 2D and $=1/2$ for 3D), increases the numerical accuracy considerably for larger amplitude $w(x)$ \cite{Zwanzig1992, Reguera2001}. The qualitative behavior of the results do not change if constant diffusivity $D_0$ is considered \cite{Malgaretti2013}. Here, for simplicity, we study our lattice model with constant diffusivity. However, it can be easily extended to incorporate x-dependent diffusivity by choosing $\lambda^{\rm ub}_{\pm}(x) = (D(x)/\ell^2) \,{\rm e}^{-\beta(\mathcal{A}(x\pm \ell) -\mathcal{A}(x))/2}$.

%\textcolor{blue}{It has been shown that the introduction of an x-dependent diffusivity, $D(x)=D_0/(1+w^{\prime}(x)^2)^{\alpha}$ (with $\alpha=1/3$ for 2D and $=1/2$ for 3D), increases numerical accuracy considerably for larger amplitude $w(x)$ \cite{Zwanzig1992, Reguera2001}. The qualitative behavior of the results do not change if constant diffusivity $D_0$ is considered \cite{Malgaretti2013}. For simplicity, we study our lattice models with constant diffusivity. However, one can easily extend our model to incorporate x-dependent diffusivity by choosing $\lambda^{\rm ub}_{\pm}(x) = (D(x)/\ell^2) \,{\rm e}^{-\beta(\mathcal{A}(x\pm \ell) -\mathcal{A}(x))/2}$.}

\subsection{Simulation Details and Parameters}

Throughout this study, the lengthscales associated with the corrugated channel are taken to be $L=1 \,{\mu m}$, $a=1/(2\pi)\,{\mu m}$, and $b=1.02/(2\pi)\,{\mu m}$. Our choice of effective widths $a$ and $b$ implies fairly strong confinement (effective bottleneck width $=2(b-a)=0.02/(2\pi) {\mu m}$). The parameter values for the two-state motion ($k_{\rm on}, k_{\rm off}, D_0, v_b,~\rm{and}~ \ell$) of the MCC are informed by experiments on kinesin motors carrying cargos or pulling membranes ~\cite{Visscher2000,Campas2008,Klumpp2005,Korn2009}.  The value of the step size or lattice spacing $\ell=8\,nm$. The simulations are performed with the free diffusion constant of the unbound MCC $D_0=0.64 \,{\mu m^2}s^{-1}$ and the off-rate $k_{\rm off}= 0.42 \,s^{-1}$, unless otherwise specified. To explore extended parameter space, the on-rate and bound velocity are varied over wide ranges, $k_{\rm on}= 0.05-50\,s^{-1}$ and  $v_b=0.04-1.6\,{\mu m}\,s^{-1}$. The experimental values of $k_{\rm on}$ ($\sim 4.7 \,s^{-1}$) and $v_b$ ($\sim0.8\,{\mu m}\,s^{-1}$) lie well within the range.

For the binding (on) rate, we study two cases: (i) $k_{\rm on}=k_{\rm on}^0$, and (ii) $k_{\rm on}(x)\propto k_{\rm on}^0/(w(x))$. In the latter case, we further investigate two situations -- when the spatial average of $k_{\rm on}$, in the interval $L$ is $k_{\rm on}^0$, to allow for comparison with (i), and when it is greater than $k_{\rm on}^0$. The simulations are performed with open boundary condition, meaning that the channel can be thought of as extending to infinity in both directions. All the data presented in this paper are averaged over $25000$ or more realizations.

\section{Results}

We characterize the motor-driven cargo transport in our model by the mean squared displacement, $\langle \delta x^2(t) \rangle=\langle (x(t+t_0)-x(t_0))^2 \rangle$, 
the average velocity, $\langle \dot{x}\rangle$, and the effective diffusivity, $D_{\rm eff}$ of the MCC. The last two quantities are defined in the asymptotic limit as $\langle \dot{x} \rangle = \lim_{t\rightarrow \infty}\frac{\langle (x(t) -x(0) \rangle}{t}$ and $D_{\rm eff} = \lim_{t\rightarrow \infty} \frac{\langle x^2(t) \rangle -  \langle x(t)\rangle^2}{2\,t}$, where $x(t)$ is the position of the particle at time $t$ and $\langle \cdot \rangle$ represents ensemble averages \cite{Reimann2002}. This definition of effective diffusivity allows for a more accurate estimate than inferring it from the MSD.  While the MSD may or may not grow linearly with time depending on context, the fluctuations around the mean position of the MCC in our systems grow linearly with time at large times and therefore $D_{\rm eff}$ is independent of time.

\subsection{Passive Transport in a Confining Channel}
We first study the passive, diffusive transport of a particle in confinement using  kinetic Monte Carlo simulations of the lattice model. We demonstrate that the lattice model correctly incorporates hopping rates through the entropic barrier dependent free energy, and discuss properties which will be used to compare and understand the results of two-state transport in the next section. In this case, a particle at the lattice site $x$ can hop to one of its neighboring sites ($x\pm \ell$) with rate $\lambda_{\pm} = (D_0/\ell^2)\,{\rm e}^{-\beta(\mathcal{A}(x\pm \ell) -\mathcal{A}(x))/2}$. For Brownian motion under constant force $F$, $\mathcal{A}(x)=-F x -T S(x)$, which gives  $\lambda_{\pm}(x) = (D_0/\ell^2)\,{\rm e}^{\pm \beta a F/2} \sqrt{w(x\pm \ell)/w(x)}$. In the absence of any confinement, the rates of hopping then become, $\lambda_{\pm}(x) = (D_0/\ell^2) \,{\rm e}^{\pm \beta \ell F/2}$, i.e. independent of $x$. Confinement makes the hopping probabilities $x$ dependent, which are given by $\lambda_{\pm}(x)/\Gamma(x)$, where $\Gamma(x) = \lambda_{+}(x) +\lambda_{-}(x)$.  

\begin{figure}[!h]
\centering
\mbox{
\hspace{-1.1cm}
\includegraphics[angle=-90,width=0.36\textwidth]{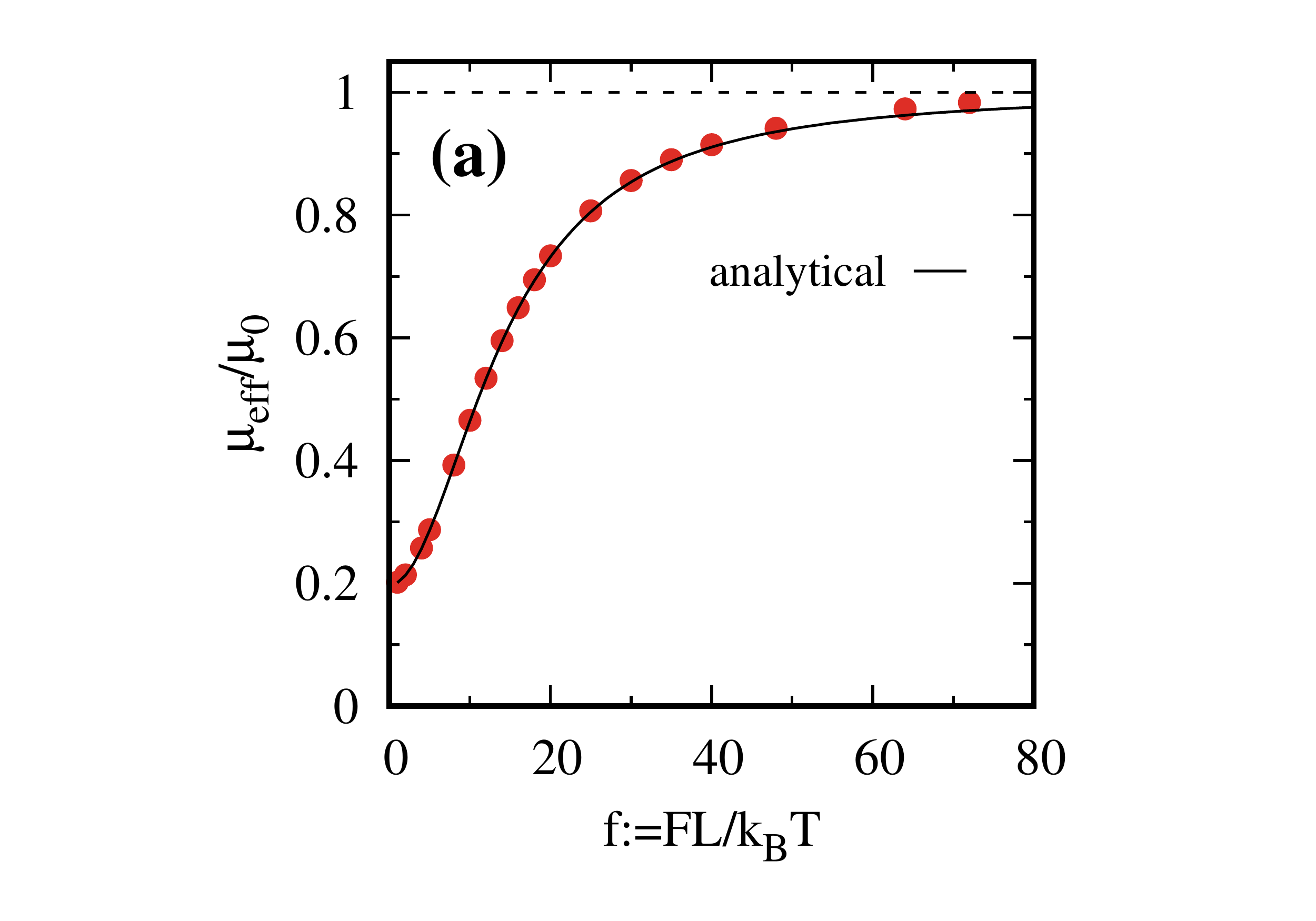}
\hspace{-2.5cm}
\includegraphics[angle=-90,width=0.36\textwidth]{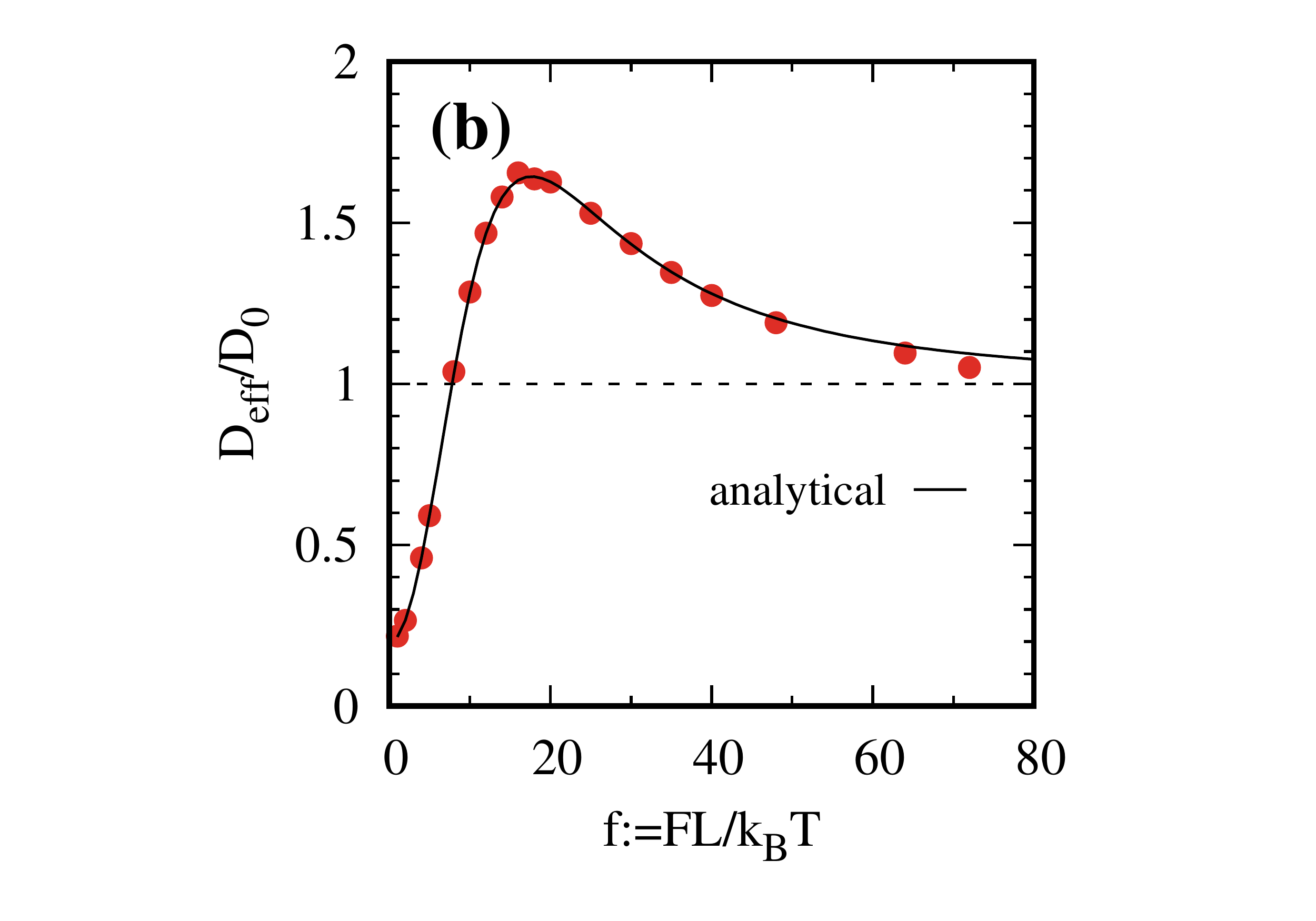}
}
\mbox{
\hspace{-1cm}
\includegraphics[angle=-90,width=0.33\textwidth]{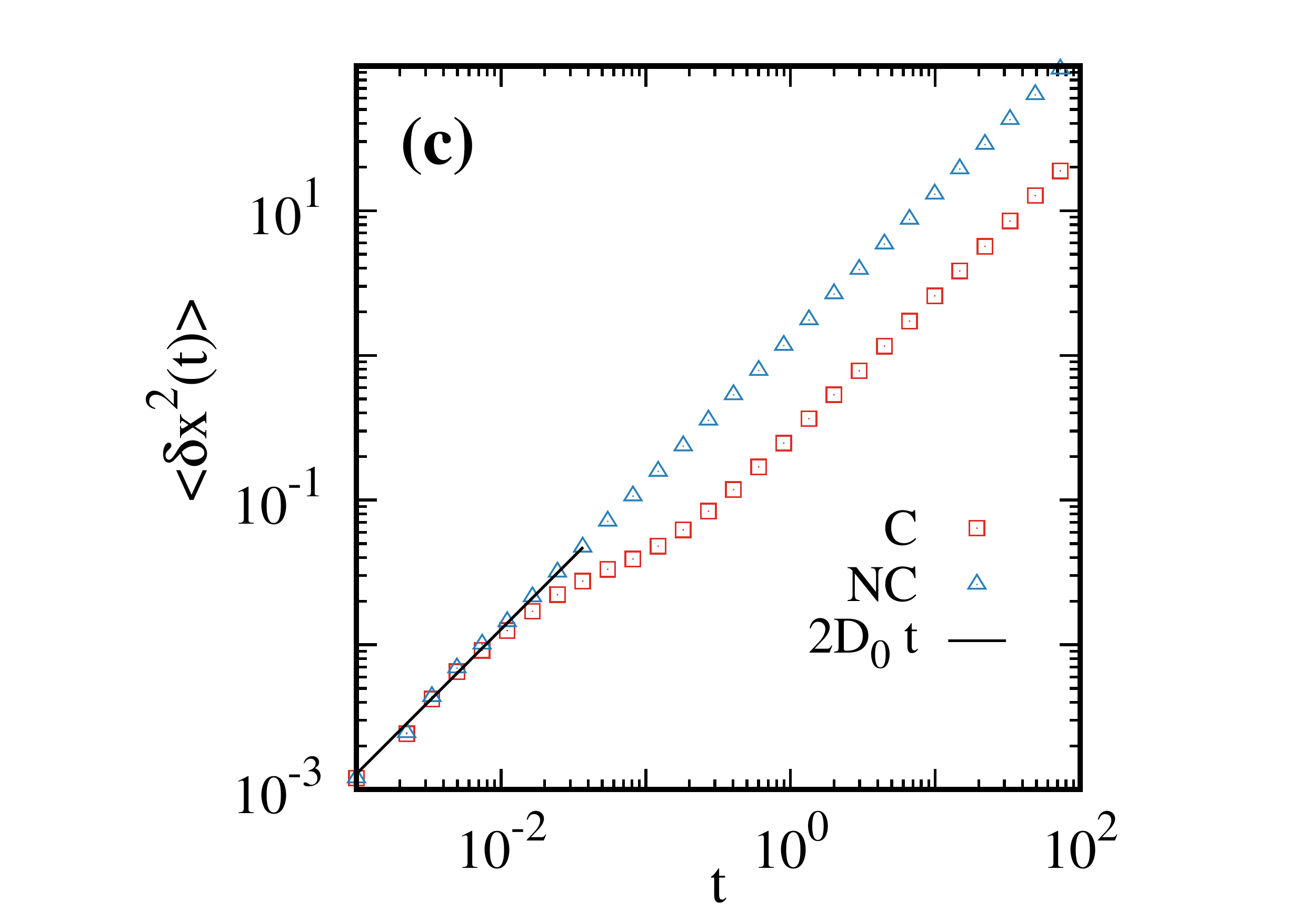}
\hspace{-1.8cm}
\includegraphics[angle=-90,width=0.33\textwidth]{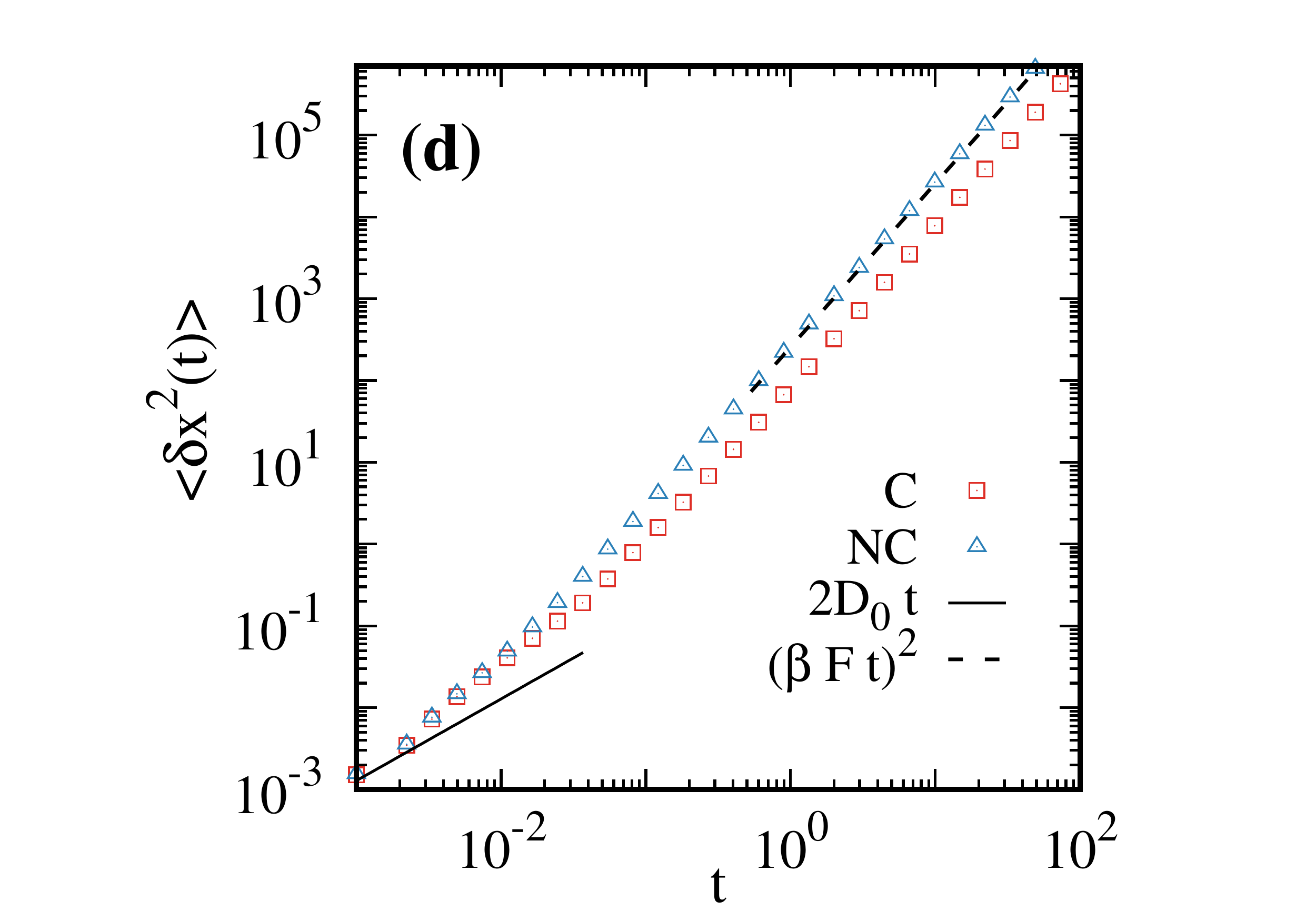}
}
\caption{Lattice model results for passive transport in a channel. Simulation data (solid circles) for (a) scaled mobility $\mu_{\rm eff}/\mu_0$ ($\mu_0=D_0/k_BT$) and (b) scaled diffusion coefficient $D_{\rm eff}/D_0$ as a function of the external driving force $f$ show very good agreement with the analytical predictions (solid lines) given by Eq.~\ref{eqn:passive}. Figure (c) shows simulation data for the MSD with time with confinement (C) and without confinement (NC) in the absence of any external driving force; Figure (d) shows the same, but with a driving force $f=25$.}
\label{fig:passive}
\end{figure}

In Fig.~\ref{fig:passive}(a) and (b), we present the lattice model results for the scaled mobility $\mu_{\rm eff}$ and scaled diffusion coefficient $D_{\rm eff}$ and compare them with the corresponding analytical predictions. The latter are obtained by solving the Fick-Jacobs equation (Eq.~\ref{eqn:fpeold}) for a particle undergoing  2D overdamped Brownian motion under an external driving force $F$ in a corrugated channel $w(x)$ with periodicity $L$~\cite{Reguera2001, Reguera2006, Burada2009,Burada2009a} and given by
\begin{subequations}
\label{eqn:passive}
 \begin{eqnarray}
\label{eqn:mu_passive}
\mu_{\rm eff} &:=& \frac{\langle \dot{x} \rangle}{F} = \frac{D_0}{k_BT}\frac{(1-\rm{e}^{-f})}{\int_0^L  \frac{dx}{L} I(x,f)} f^{-1}, \,\textrm{and}\\\nonumber 
\label{eqn:d_passive}
\frac{D_{\rm eff}}{D_0} &=& \int_0^L \frac{dx}{L} \int_{x-L}^x \frac{dz}{L} \frac{{\rm e}^{\mathcal{A}(x)/k_BT}}{{\rm e}^{\mathcal{A}(z)/k_BT}} I^2(z,f)\\
&&\times\left[ \int_0^L I(x,f) \frac{dx}{L} \right]^{-3},
\end{eqnarray}
\end{subequations}
where
 $I(x,f) := {\rm e}^{\mathcal{A}(x)/k_BT} \int_{x-L}^x \frac{dy}{L} {\rm e}^{-\mathcal{A}(y)/k_BT}$ depends on the dimensionless force (or Peclet number) $f:=FL/k_BT$. Please note that the above expressions are nonlinear in $f$. In the absence of any geometric confinement, the effective mobility and diffusion coefficient reduce to $\mu_{\rm eff}=D_0/k_BT=\mu_0$ and $D_{\rm eff}=D_0$, by substituting $\mathcal{A}(x)=-Fx$ in the Eq.~\ref{eqn:passive}. The numerical results are in very good agreement with the analytical predictions, demonstrating that our lattice model is an accurate representation of the Fick-Jacobs equation (Eq.\ref{eqn:fpeold}), and suggesting that this method can be used to study  a wide range of systems with entropic barriers.  
 
We now discuss the results shown in Fig.~\ref{fig:passive} in more detail. The scaled mobility shown in Fig.~\ref{fig:passive}(a) is always less than 1,  approaching 1 asymptotically as $f$ is increased. This suggests that a symmetric confinement without any rectification mechanism cannot enhance the mobility of a purely diffusive system. The behavior of the scaled effective diffusivity $D_{\rm eff}/D_0$ in Fig.~\ref{fig:passive}(b) is non-monotonic with a peak at a critical value of $f$, suggesting that while at small $f$ confinement causes the effective diffusivity to decrease, at large $f$ the interplay of the force and confinement leads to enhanced diffusivity. The value of the critical force depends on the modulation of the confining wall as discussed in ref \cite{Burada2008}. 

%In Fig.~\ref{fig:passive}(c) and (d), we show the MSD for $f=0$ and $f=25$ respectively, and compare the results with the corresponding unconfined case. In the absence of confinement, the motion of the particle is determined by the dimensionless quantity $f$ which describes the relative propensity for ballistic vs diffusive motion, or in order words is the Peclet number for this case. For $f=0$ (Fig.~\ref{fig:passive} (c)), the particle's motion is predominantly diffusive, while for large $f$ (Fig.~\ref{fig:passive}(d)), it is largely ballistic. 

Confinement constrains diffusive motion of the particle by reducing the available space for movement; for $f=0$, it leads to a significant decrease in the MSD compared to the unconfined case at intermediate and large times as seen in Fig.~\ref{fig:passive}(c); the particle's motion changes from free diffusion with $D_0$ at early times to an intermediate sub-diffusive regime, and finally to effective diffusive behavior with $D_{\rm eff}<D_0$ at large times. The intermediate sub-diffusive regime presumably emerges due to the slowing down of motion near the neck of the channel. For large Peclet numbers $f>>1$, (Fig.~\ref{fig:passive} (d)), where the particle motion is largely driven rather than diffusive, confinement, has much smaller impact compared to (c). The motion changes from free diffusion at very early times to force-driven ballistic motion at large times, and as in (c) there is a visible slowing down in at intermediate times for the confined case, but the gap between the two asymptotic MSDs is much smaller. As we will see in the next section, the intermediate slowing down due to confinement plays a critical role in two-state cargo transport.

\subsection{Active and Passive Transport in a Confining Channel}
\label{sec:two-state}

\begin{figure}[!h]
\centering
\mbox{
\hspace{-1.1cm}
\includegraphics[angle=-90,width=0.35\textwidth]{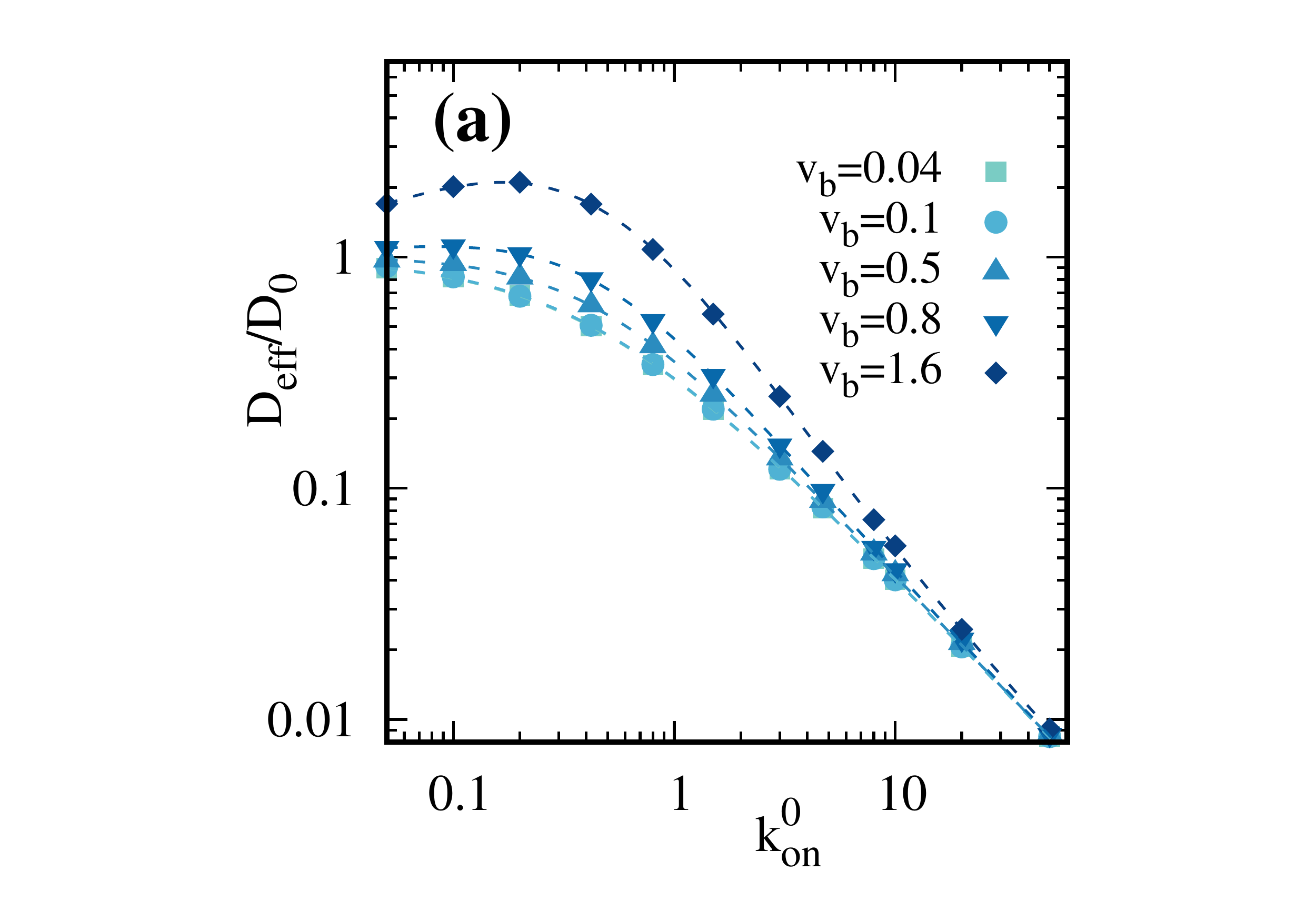}
\hspace{-2.2cm}
\includegraphics[angle=-90,width=0.35\textwidth]{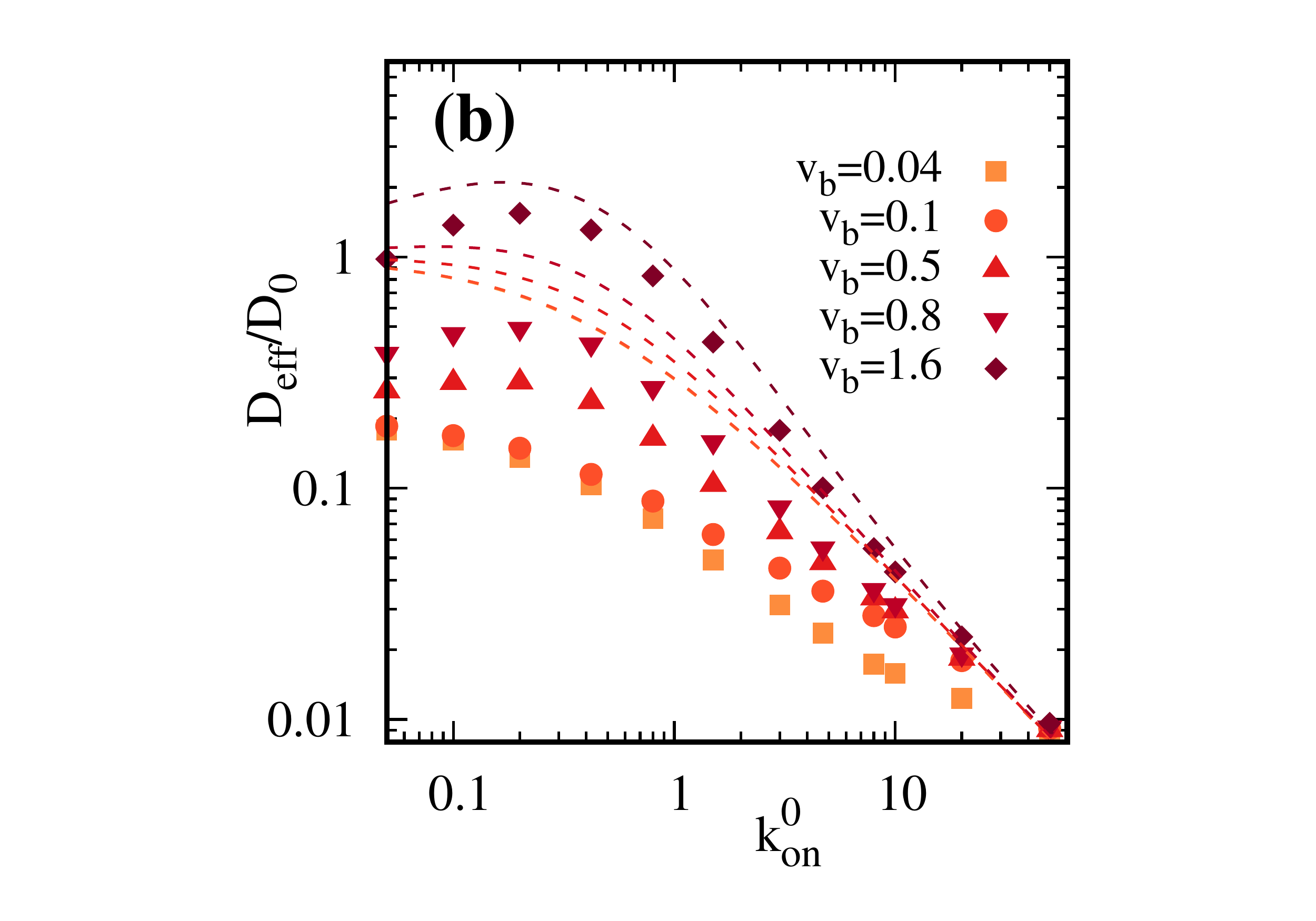}
}
\caption{Scaled effective diffusivity $D_{\rm eff}/D_0$ as a function of the binding rate $k_{\rm on}$ (a) without confinement and (b) with confinement for various values of $v_b$.
We have set $k_{\rm off}=0.42\,s^{-1}$, and $D_0 = 0.64 \,{\mu m^2} s^{-1}$. Dashed lines represent the MFNC limit. The data for unconfined case follow the MFNC limit, while for confined case data stay below their MFNC limit.}
\label{fig:active2}
\end{figure}

We now discuss the transport properties, namely MSD, average velocity, and effective diffusivity of the MCC for the two state model with confinement, and compare them with the results for the unconfined case. 
Where appropriate, we also compare our results with corresponding steady state values in the mean field limit without any confinement. The following sections we refer to this limit as the Mean Field No Confinement (MFNC) limit. 
In this limit, the probabilities of bound and unbound states are given by $\tilde{P}_{b}=k_{\rm on}/(k_{\rm on}+k_{\rm off})$ and $\tilde{P}_{ub}=k_{\rm off}/(k_{\rm on}+k_{\rm off})$, respectively, and the average velocity and effective diffusivity can be written as,
\begin{eqnarray}
\label{eqn:limitv}
V_l &=& \tilde{P}_b v_b = \frac{k_{\rm on}v_b}{k_{\rm on} + k_{\rm off}},\\
\label{eqn:limitd}
D_l &=& \tilde{P}_{ub} D_0 + D_{act,ub} = \frac{k_{\rm off}D_0}{k_{\rm on} + k_{\rm off}} + \frac{k_{\rm on} k_{\rm off} v_b^2}{(k_{\rm on} + k_{\rm off})^3}. 
\end{eqnarray} 
In Eq.~\ref{eqn:limitv}, the velocity $V_l$ does not depend on the probability of the unbound state since there is no net directed movement during passive motion. In Eq.~\ref{eqn:limitd}, the first term is due to the diffusive motion in the unbound state and the second term accounts for an additional contribution due to the stochastic transition between active and passive motion. The analytical expression for the latter can be found in \cite{Dogterom93}. Maximizing this equation provides the condition for the occurrence of the peaks in effective diffusivity as well as their positions, as observed in (Fig.~\ref{fig:active2}(a)). 

The above mean field analytical expressions are exact for the unconfined case if the transition rates and the diffusivity of the motion are independent of position. As the confinement makes the diffusivity position dependent (through position dependent hopping rates), these expressions are no longer expected to hold true for the confined case. However, comparing results with these expressions are very useful for understanding the role of the confinement.

\begin{figure}[!h]
\centering
\mbox{
\hspace{-1cm}
\includegraphics[angle=-90,width=0.32\textwidth]{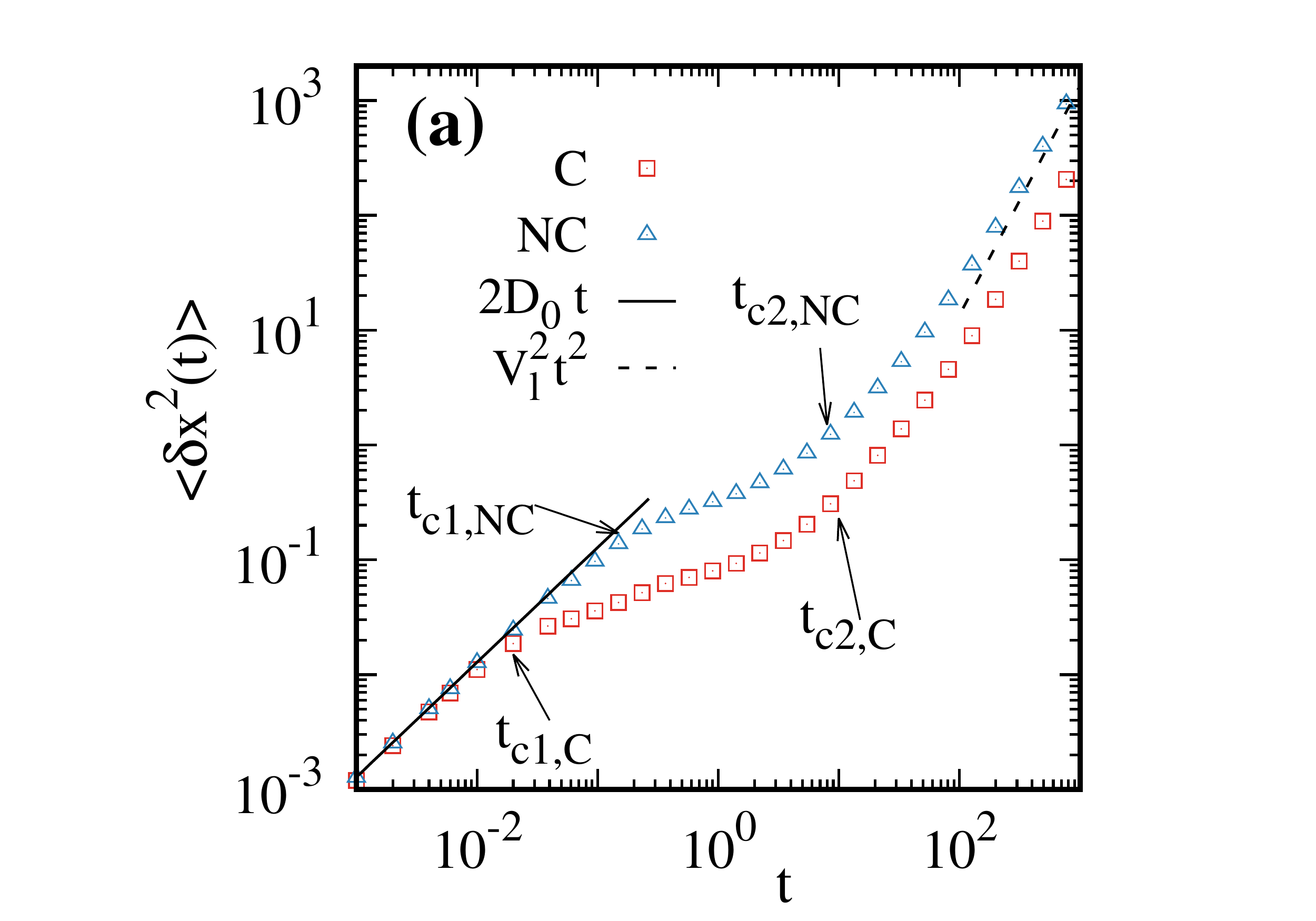}
\hspace{-1.5cm}
\includegraphics[angle=-90,width=0.32\textwidth]{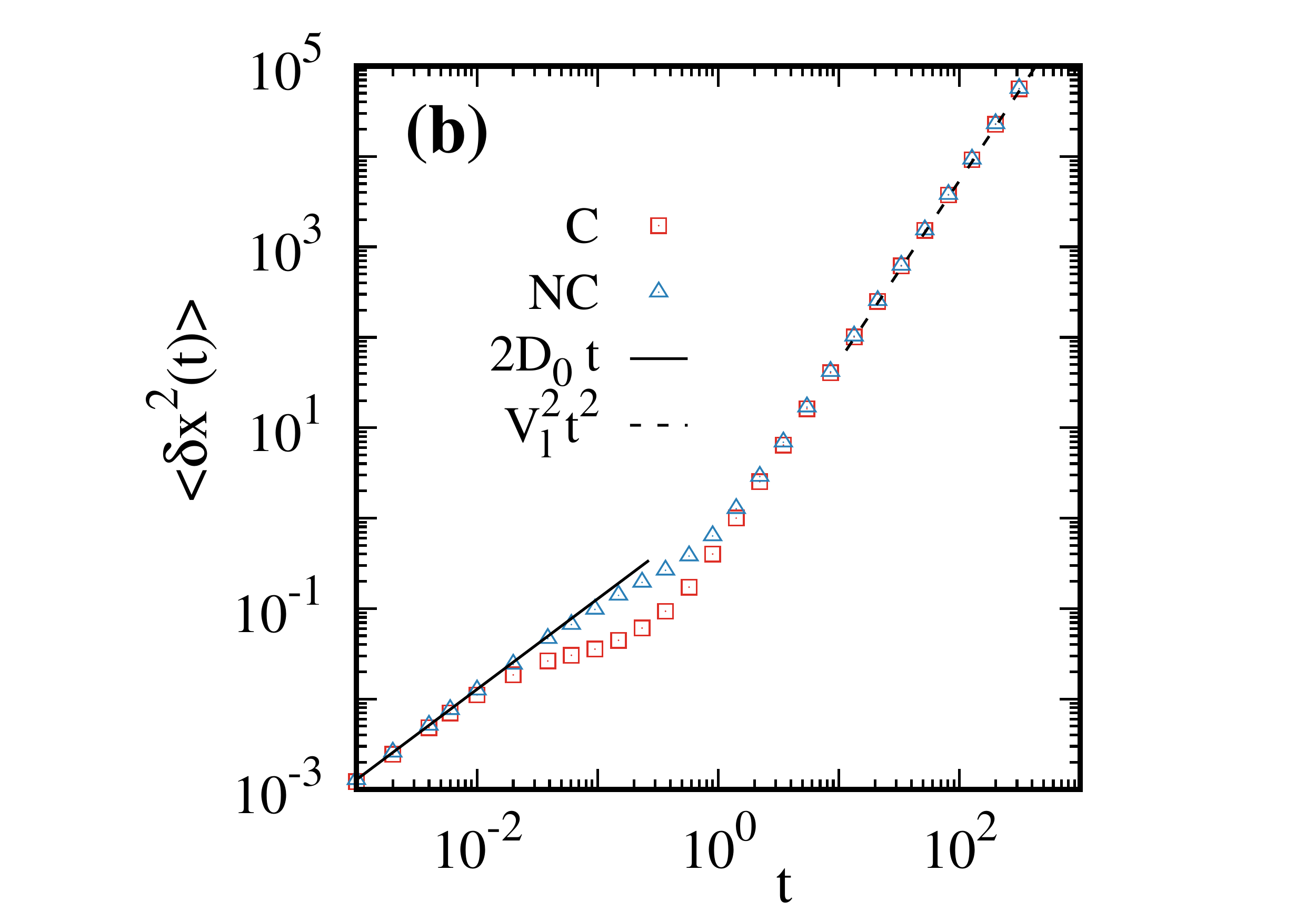}
}
\mbox{
\hspace{-1cm}
\includegraphics[angle=-90,width=0.32\textwidth]{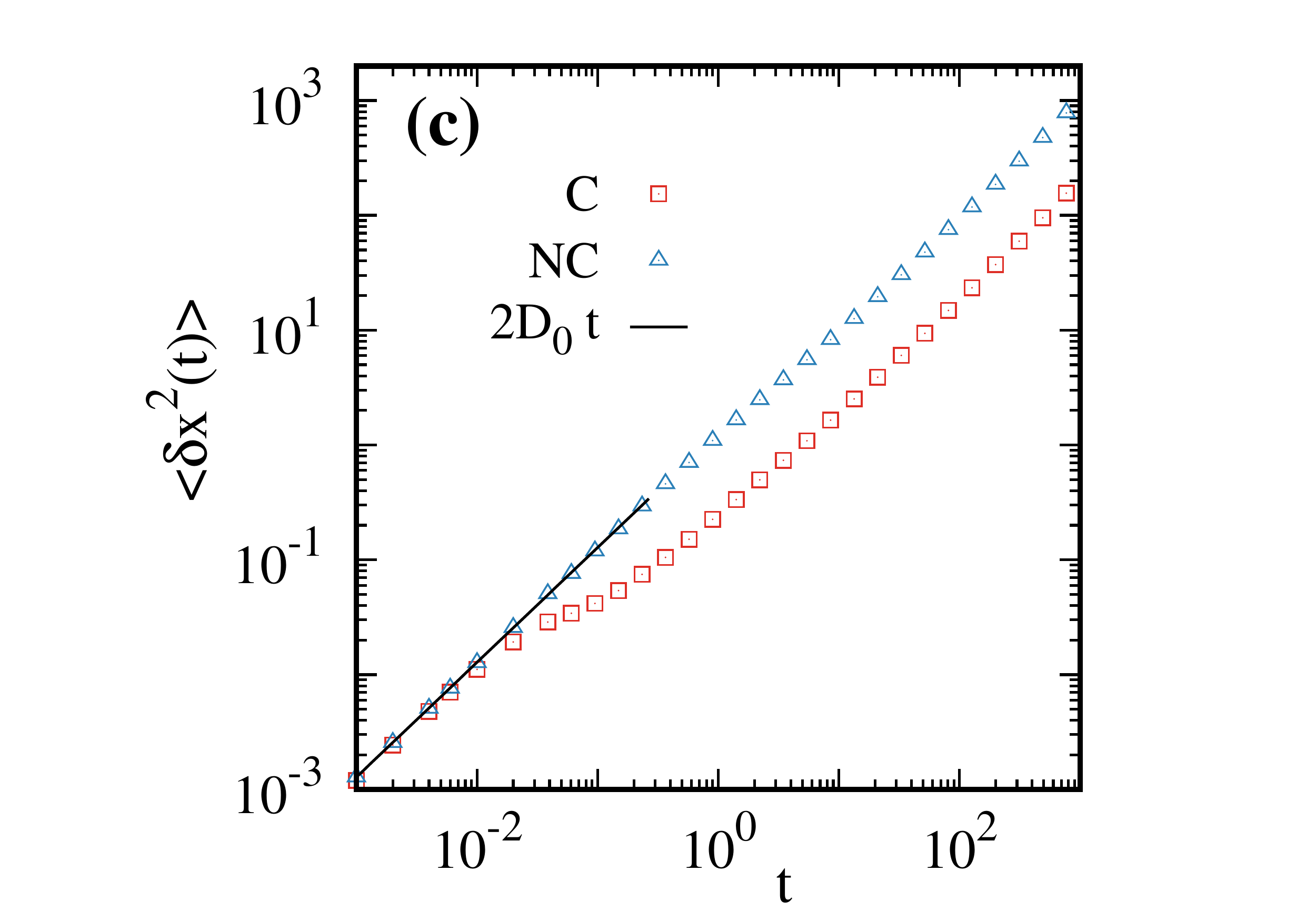}
\hspace{-1.5cm}
\includegraphics[angle=-90,width=0.32\textwidth]{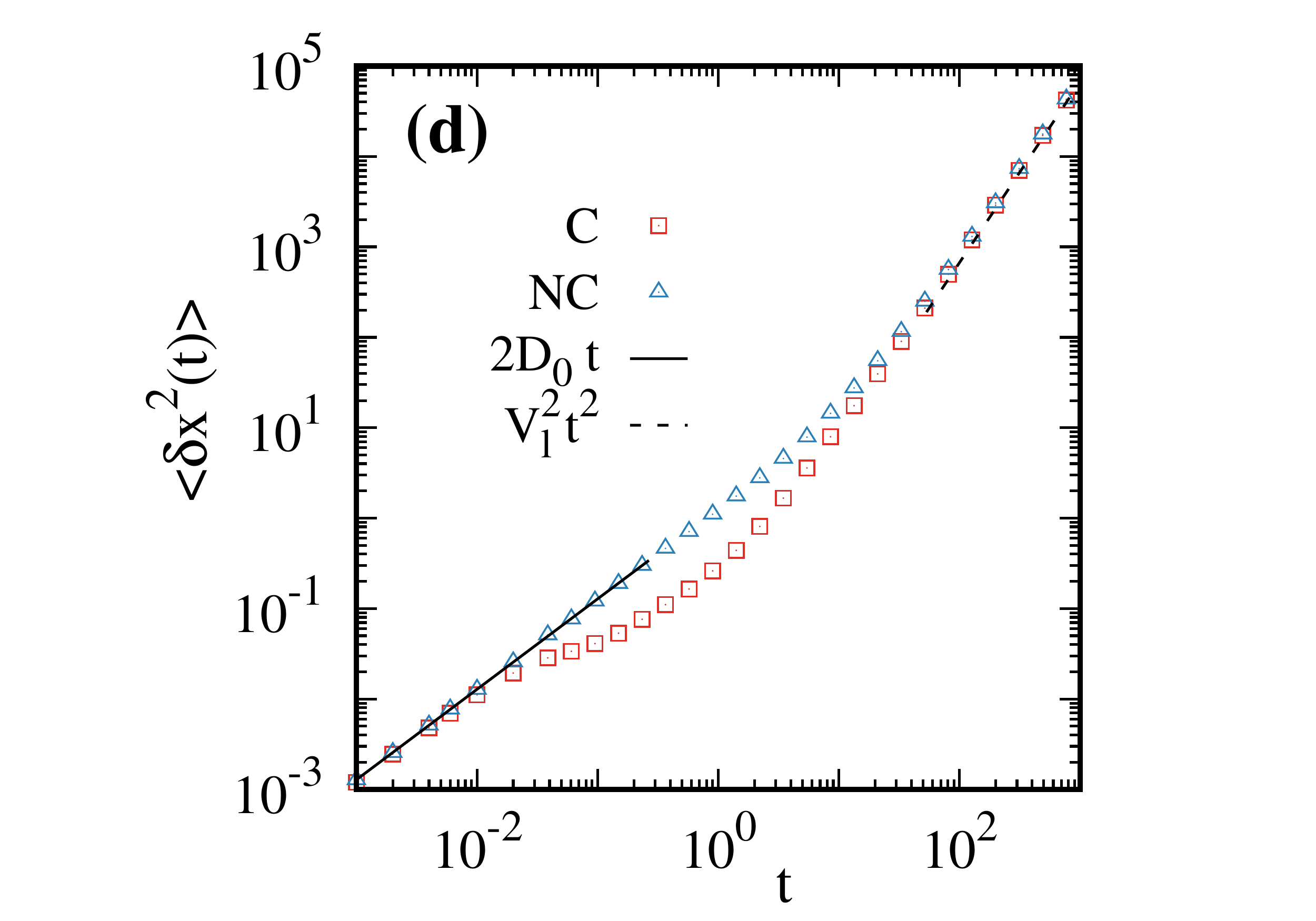}

}
\caption{MSD with time for $k_{\rm off}=0.42\,s^{-1}$, and $D_0=0.64 \,{\mu m^2}s^{-1}$, with confinement (C) and without confinement (NC). Figures (a) \& (b) show data for a large binding rate ($k_{\rm on}=4.7 \,s^{-1}$) of the motor to the microtubule. (a) For a small bound velocity ($v_b=0.04\,{\mu m}\,s^{-1}$), the MSDs for both C and NC grow as $2 D_0 t$ initially, deviating at intermediate times, and asymptotically converging to ballistic motion $\simeq \langle \dot{x} \rangle^2t^2$ with $\langle \dot{x} \rangle_{NC}=V_l>\langle \dot{x} \rangle_C$ at large times. (b) For a large bound velocity ($v_b=0.8\,{\mu m}\,s^{-1}$), the small $t$ diffusive motion and large $t$ ballistic motion are the same for both C and NC, but the intermediate behavior are different. Figures (c) \& (d) show data for a small binding rate ($k_{\rm on}=0.2 \,s^{-1}$), for a small velocity ($v_b=0.04\,{\mu m}\,s^{-1}$) in (c) and a large velocity ($v_b=0.8\,{\mu m}\,s^{-1}$) in (d).}
\label{fig:msd}
\end{figure}

{\bf Mean Squared Displacements (MSD):} In Fig.~\ref{fig:msd}, we present the MSD for small and large bound velocities $v_b$ for two different binding rates  $k_{\rm on}$. At very small time $t$, the MSD behaves as $\langle \delta x^2(t) \rangle \simeq 2 D_0 t$. This behavior suggests that below a crossover time scale, say $t_{c1}$, the effect of binding/unbinding kinetics and confinement are negligible such that the MCC can diffuse freely. For $t>t_{c1}$, it shows a transition from free diffusion to ``sub-diffusion''. In the absence of confinement, the sub-diffusive behavior is due to the time spent by the MCC alternately transitioning between bound and unbound states, and the crossover time scale, $t_{c1}$, mainly depends on the transition rates. In the confined case, in addition to transition events, the motion of the MCC slows down even further due to the strong entropic barrier in the diffusive state close to a neck of the channel. Consequently, the crossover time $t_{c1}$ becomes even smaller for the confined case than that without any confinement. There is a second crossover from sub-diffusive to ballistic behavior i.e., for $t>t_{c2}$, $ \langle \delta x^2(t) \rangle \simeq  \langle \dot{x} \rangle^2 t^2$.  A representative case of the locations of $t_{c1}$ and $t_{c2}$ are shown in Fig.~\ref{fig:msd} (a). From the Fig.~\ref{fig:msd}, it is clear that confinement impacts the motion of the MCC more strongly for small $v_b$ -- it spends longer times in the intermediate sub-diffusive regime leading to smaller asymptotic velocities at large times compared to the unconfined case.  For larger $v_b$, the sub-diffusion regime shrinks and the MCC has the same asymptotic velocities with and without confinement. The observed MSDs in Fig.~\ref{fig:msd} (c) and (d) are similar to (a) and (b) respectively, but with the crossover to ballistic motion occurring at much larger times ($t_{c2}$).

The motion of the MCC and how its MSD will scale with time is determined by the interplay of four timescales: the inverse of the binding rate $k_{\rm on}$, the inverse of the unbinding rate $k_{\rm off}$, the diffusion time $\tilde{L}^2/D_0$, and the drift timescale $\tilde{L}/v_b$, where $\tilde{L}$ is the characteristic length scale in the system and is equal to $L$ in the confined case. The precise dependence of $t_{c1}$ and $t_{c2}$ on these timescales is nontrivial and will be studied in future work. Here we make the following qualitative observations for our confining channel. For a given $k_{\rm off}$, the value of the first crossover timescale $t_{c1,C}$ seems to scale as $D_0/L^2$ and seems independent of $v_b$ and $k_{\rm on}$, while the second crossover timescale $t_{c2,C}$ decreases with increasing $v_b$ and $k_{\rm on}$.

\begin{figure}[!h]
\centering
\mbox{
\hspace{-1cm}
\includegraphics[angle=-90,width=0.35\textwidth]{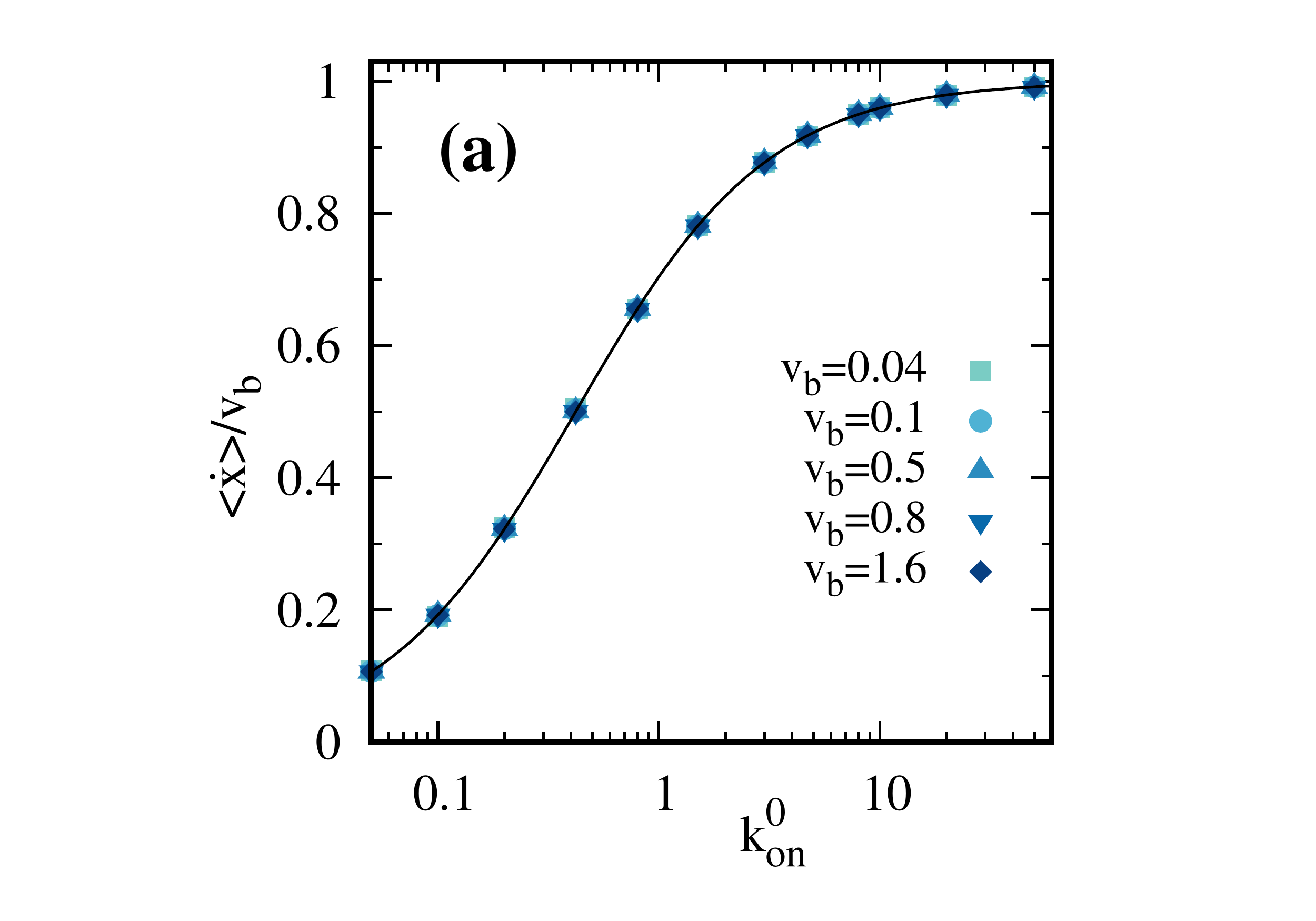}
\hspace{-2cm}
\includegraphics[angle=-90,width=0.35\textwidth]{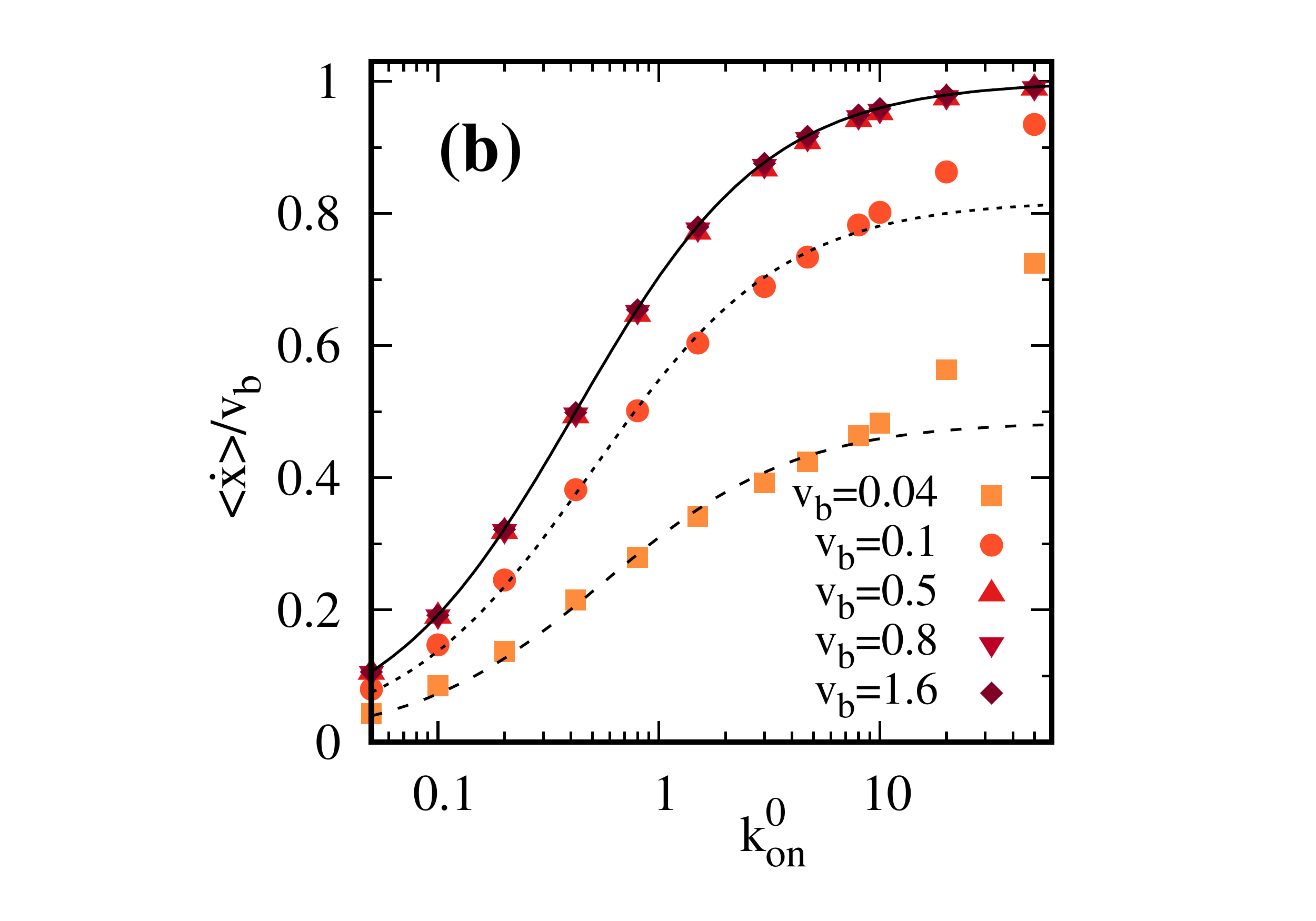}
}
\caption{Scaled average velocity $\langle \dot{x} \rangle/v_b$ of the MCC as a function of the binding rate $k_{\rm on}$ (a) without confinement and (b) with confinement for various values of $v_b$. We have set $k_{\rm off}=0.42\,s^{-1}$, and $D_0 = 0.64 \,{\mu m^2}s^{-1}$. Solid lines represent MFNC limit. The data for $v_b=0.04$, and $0.1 \, \mu m\,s^{-1}$ do not follow MFNC limit. The dashed line is a fit of $0.04 \, \mu m\,s^{-1}$ data to Eq.~\ref{eqn:limitv} which yields $k^{\rm eff}_{\rm off}= 0.57 \,s^{-1}$ and $v^{\rm eff}_b=0.019\, \mu m\,s^{-1}$. The dotted line is a fit of $0.1 \, \mu m\,s^{-1}$ data with $k^{\rm eff}_{\rm off}= 0.49 \,s^{-1}$ and $v^{\rm eff}_b=0.08\, \mu m\,s^{-1}$.}
\label{fig:active1}
\end{figure}

{\bf Average velocity:}  In Fig.~\ref{fig:active1}(a) and (b), we show the results for the scaled average velocity for various values of $v_b$, without and with confinement, respectively. 
In the absence of confinement, for large separation between the times spent by the MCC in the unbound and bound states (Eq.~\ref{eqn:limitv}), the average velocity approaches constant values-- for $k_{\rm on} \gg k_{\rm off}$,  $\langle \dot{x} \rangle$ approaches  $v_b$ and motion of the MCC is predominantly ballistic, while for for $k_{\rm on} \ll k_{\rm off}$, it approaches zero suggesting diffusive (or no) motion. In the intermediate regime, their is an interplay between ballistic and diffusive motion, and the scaled average velocities grow monotonically with $k_{\rm on}$. Fig.~\ref{fig:active1} (a) captures this behavior, with the data for all $v_b$ following Eq.~\ref{eqn:limitv}, as expected. 

Confinement reduces the average velocity of the MCC for small bound velocities $v_b$, while for large $v_b$ the results follow the MFNC limit (Fig.~\ref{fig:active1}(b)). For small $v_b$ ($0.04$, and $0.1  \mu m\,s^{-1}$), confinement leads to a reduction in the scaled average velocity and it stays below the MFNC limit described by Eq.~\ref{eqn:limitv}. However, the data can be fit to Eq.~\ref{eqn:limitv} with an effective $k^{\rm eff}_{\rm off}$, and an effective $v^{\rm eff}_b$. We find that $k^{\rm eff}_{\rm off}>k_{\rm off}$ and $v^{\rm eff}_b < v_b$ suggesting that confinement renormalizes the bound velocity and the unbinding rate to values lower than without confinement, and therefore effectively reduces the processivity of the motor. For large $v_b (0.5, 0.8,\textrm{and}, 1.6\, \mu m\,s^{-1})$, the scaled average velocities are unaffected by confinement. These results are consistent with MSDs discussed earlier.

{\bf Effective Diffusivity :} Confinement has a much more striking impact on the effective diffusivity of the MCC. Unlike for the average velocity, we do not find a data collapse in this case.  We show the scaled diffusion coefficient as a function of $k_{\rm on}$ for the unconfined case in Fig.~\ref{fig:active2}(a). For small $k_{\rm on}$, the motion of the MCC is dominated by passive diffusion and $D_{\rm eff}$ is large.  While for large $k_{\rm on}$, as the motion is largely ballistic,  $D_{\rm eff}$ become smaller. As expected, the data for the unconfined case follow the analytical prediction given by Eq.~\ref{eqn:limitd}. The condition for a maximum to exist is predicted by Eq. \ref{eqn:limitd}: $v_b > \sqrt{k_{\rm off}/D_0} (= 0.81$ for parameters used here). We observed that the scaled diffusion coefficient shows a maximum for $v_b=1.6\,\mu m\,s^{-1}$ at $k_{\rm on} \simeq 0.2\,s^{-1}$. In the case of confinement (Fig.~\ref{fig:active2}(b)), the effective diffusion coefficient becomes smaller than for the case without confinement, but they are well separated with prominent peaks, and the presence of a maximum at $k_{\rm on}\simeq0.2\,s^{-1}$ is now seen for small $v_b$. For $v_b=0.8\,\mu m\,s^{-1}$ we observe a peak here which is not possible for the unconfined case. The lowering of effective diffusivity suggests that the confinement reduces the noise in the motion by limiting the availability of space for the motion. 

%We observed that scaled diffusion coefficient shows a maximum for $v_b=1.6\,\mu m\,s^{-1}$ at $k_{\rm on} \simeq 0.2\,s^{-1}$. The condition for such a maximum to exist is predicted by Eq. \ref{eqn:limitd}: $v_b > \sqrt{k_{\rm off}/D_0} (= 0.81$ for parameters used here). 
%In the case of confinement (Fig.~\ref{fig:active2}(b)), the effective diffusion coefficient becomes smaller than for the case without confinement, but they are well separated with prominent peaks, and the presence of a maximum at $k_{\rm on}=0.2\,s^{-1}$ is now seen for small $v_b$. 

The observed maximum in diffusivity is reminiscent of similar behavior for a particle undergoing diffusion in a tiled washboard potential~\cite{Costantini1999}, or  in a periodic confined profile~\cite{Reguera2006}. There the appearance of a peak in the diffusivity is associated with a ``locked-to-running'' transition~\cite{Costantini1999}. In the locked state, the particle shows no net movement over a significant amount of time, while in the running state the particle has a net drift velocity. Transitions between the two states can be induced via occasional large kicks due to noise.  We observe peaks in the two-state model even without any confinement, with the passive (unbound) and active (bound) states corresponding to the locked and running states respectively. The transition between the two states are induced by the binding kinetics of the motors to the microtubule. Confinement makes the peaks much more pronounced, and leads to greater separation between the scaled effective diffusivity curves for different $v_b$. 

The qualitative behavior of the average velocity and effective diffusivity does not depend on the free diffusion constant $D_0$ and the effective widths of the channel. We have checked this by studying systems with $D_0=0.064 \,\mu m^2 s^{-1}$ and $b=1.2/(2\pi) \mu m$ (results not presented here).

\subsection{Active and Passive Transport in a Confining Channel with Spatially Varying Binding Rates}

Next we study cargo transport by motors with a spatially varying binding rate that depends on the local width of the confining channel.  
 We have studied the following two cases: (i) $k_{\rm on}(x) = k^0_{\rm on}\sqrt{b^2-a^2}/(L\,w(x))$, where the binding rate is normalized to ensure that the spatial average $\langle k_{\rm on}(x) \rangle= k^0_{\rm on}$, and  (ii) $k_{\rm on}(x) = k^0_{\rm on}/w(x)$, where the binding rate is not normalized, and its spatial average  $\langle k_{\rm on}(x) \rangle= k^0_{\rm on}L/\sqrt{b^2-a^2}$. For both (i) and (ii),  the unbinding rate $k_{\rm off}$ is assumed to be independent of the spatial variation of the channel width.

\begin{figure}[]
\centering
\mbox{
\hspace{-1cm}
\includegraphics[angle=-90,width=0.35\textwidth]{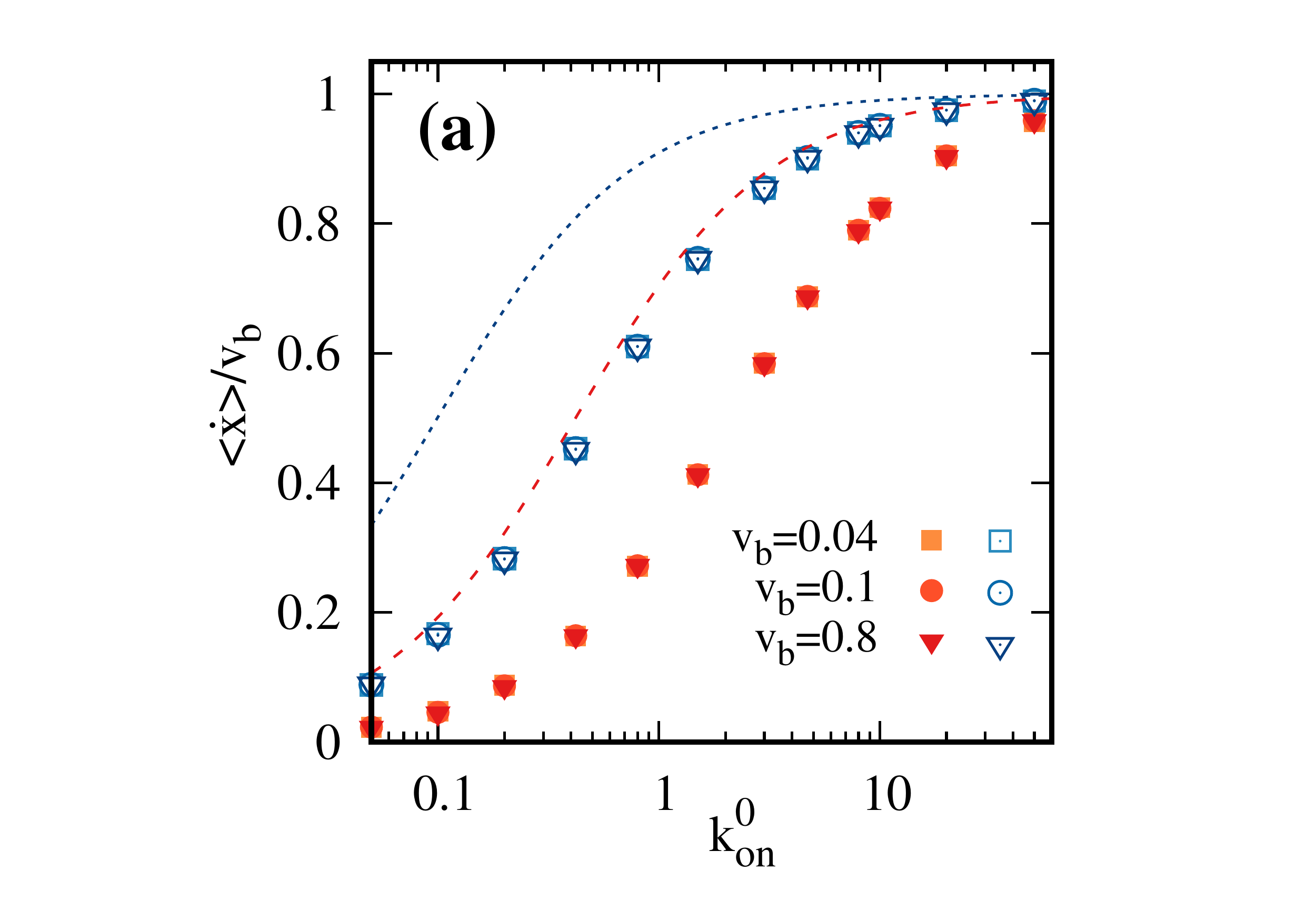}
\hspace{-2cm}
\includegraphics[angle=-90,width=0.35\textwidth]{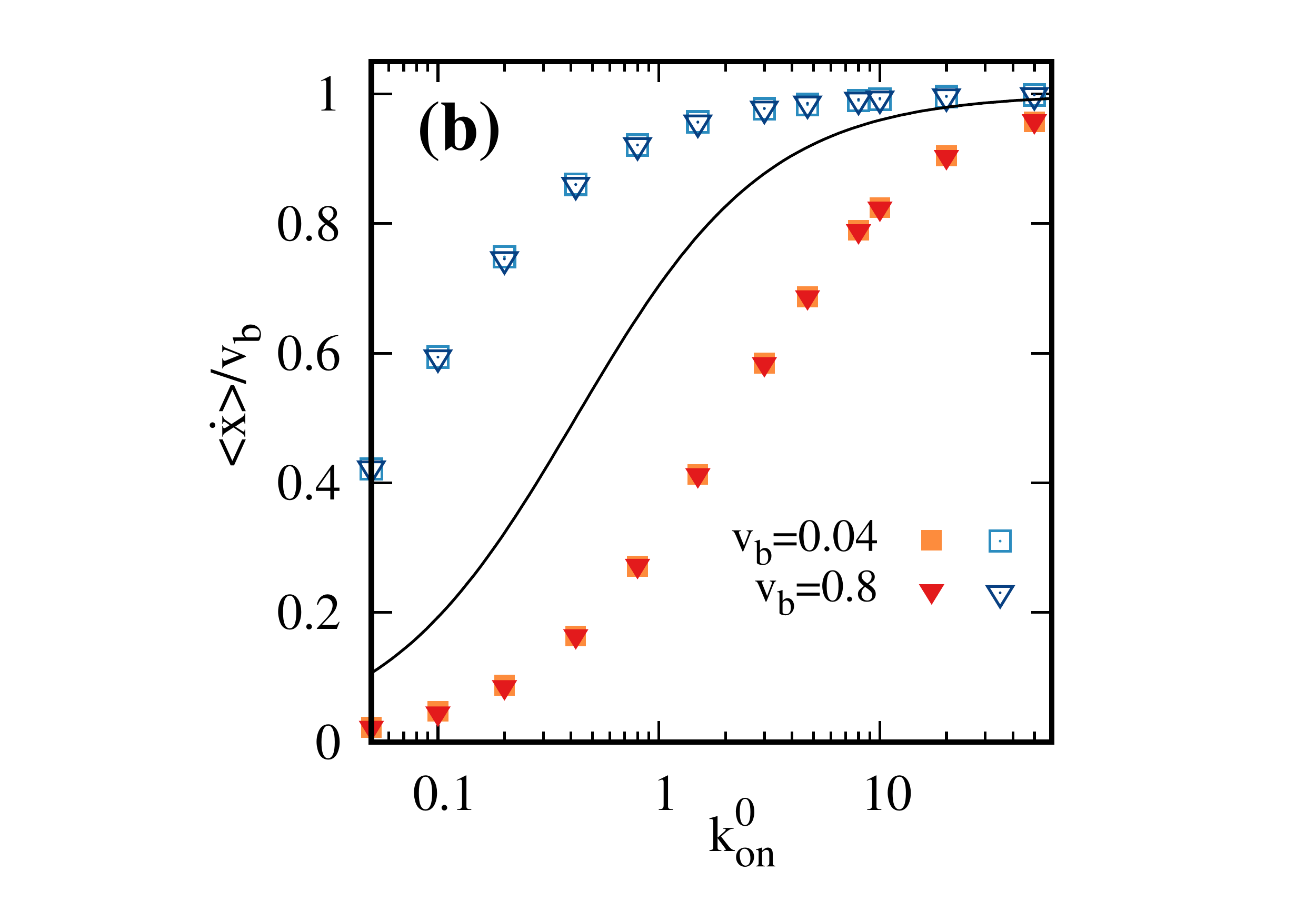}
}
\mbox{
%\hspace{-1cm}
\includegraphics[angle=-90,width=0.35\textwidth]{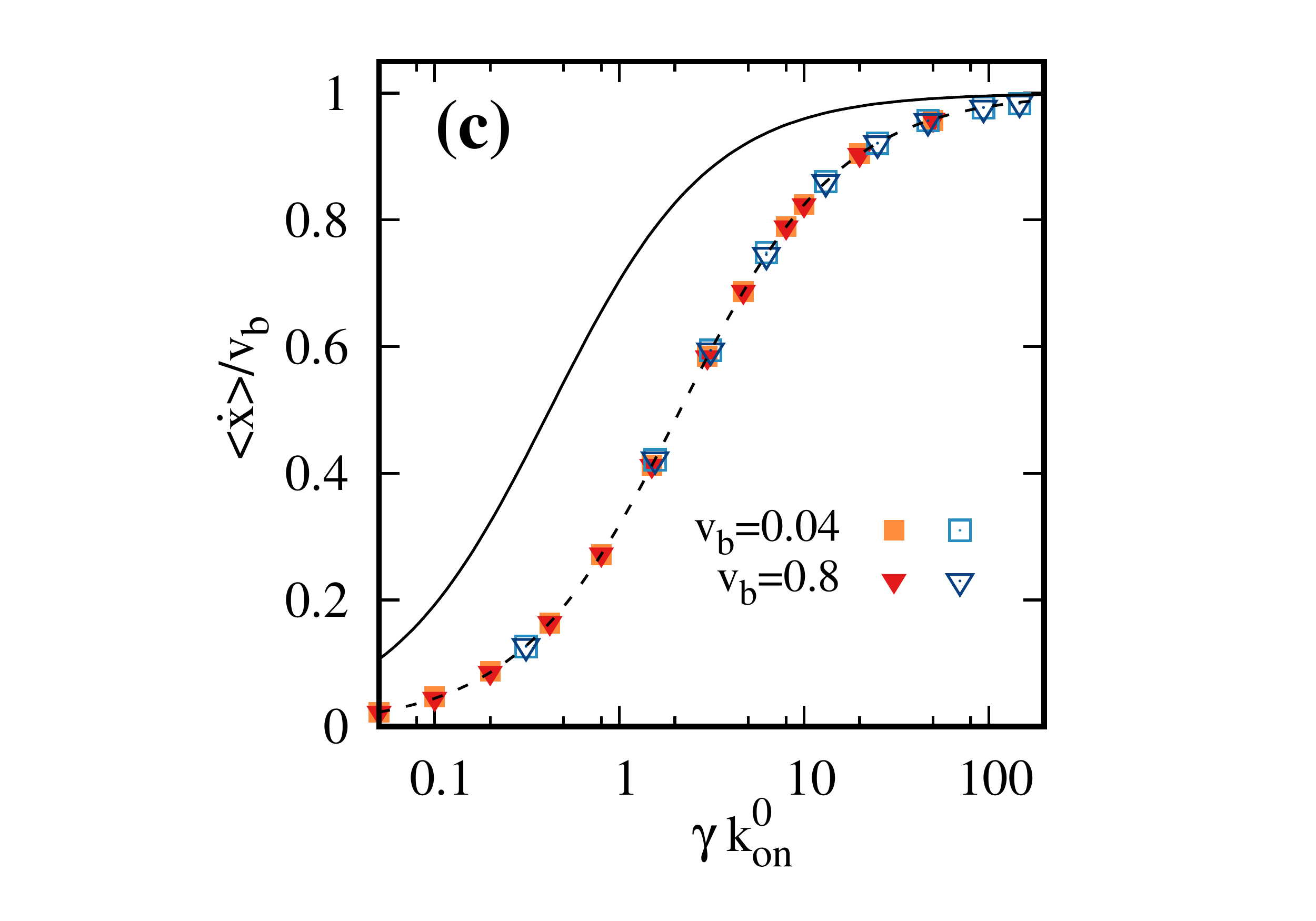}
}
\caption{Scaled average velocity as a function of $k^0_{\rm on}$ for motors  with spatially varying binding rate $k_{\rm on}(x) = k^0_{\rm on}/w(x)$.  (a) Shows data for case (i)  with $k_{\rm off}=0.42 \,s^{-1}$ (solid symbols) and $k_{\rm off}=0.0955\,s^{-1}$ (open symbols) for $v_b=0.04, 0.1,$ and $0.8 \,\mu m \, s^{-1}$. The corresponding MFNC predictions are shown with dotted and dashed lines respectively. (b) Shows data for case (i) (solid symbols) and case (ii) (open symbols) for $k_{\rm off}=0.42 \, s^{-1}$, and the MFNC prediction (solid line). (c) Shows the data collapse as a function of $\gamma k^0_{\rm on}$, where $\gamma=1$ for case (i) and $\gamma=L/\sqrt{b^2-a^2}$ for case (ii). The solid line represents the MFNC limit for $k_{\rm off}=0.42 \, s^{-1}$, and the dashed line represents the fit to MFNC limit with an effective unbinding rate $k^{\rm eff}_{\rm off}=2.14 \, s^{-1}$. Note that $k^{\rm eff}_{\rm off}$ is very large compared to the actual unbinding rate.}
\label{fig:active_np}
\end{figure}

Let us first consider the case (i) with the normalized spatially varying binding rate. We present the scaled velocities against $k^0_{\rm on}$ for two different values of unbinding rates $k_{\rm off}=0.42$ and $0.0955 \,s^{-1}$ in Fig.~\ref{fig:active_np}(a). For each value of the unbinding rate, we consider three different bound velocities $v_b$. For each unbinding rate, the average velocity data collapse onto a single curve, suggesting that unlike for constant binding rates the scaled velocities do not depend on $v_b$. We further find that for both unbinding rates, the average velocities of the MCC stay below the corresponding velocities $V_l$  in the MFNC limit given by Eq.~\ref{eqn:limitv}. The decrease in the scaled velocity suggests that the effective unbinding rate $k^{\rm eff}_{\rm off}$ has increased, and the collapse to a single curve suggests that $k^{\rm eff}_{\rm off}$ is independent of bound velocity $v_b$ (see Fig.~\ref{fig:active_np}(c)), unlike the case with constant binding rate.

Next we compare case (i) with case (ii), where the binding rate is not normalized so that the average $\langle k_{\rm on}(x) \rangle$ is not equal to $k^0_{\rm on}$ and is, in fact, larger than $k^0_{\rm on}$. 
In Fig.~\ref{fig:active_np}(b), we show the scaled velocities against $k^0_{\rm on}$ for case (i) and case (ii), for a given unbinding rate $k_{\rm off}=0.42\,s^{-1}$ along with the corresponding MFNC prediction (Eq.~\ref{eqn:limitv}). We observe that for both cases, the data for all bound velocities show a good collapse as in Fig.~\ref{fig:active_np}(a). More interestingly, the average velocity of the MCC for case (ii)  
is greater than the MFNC velocity $V_l$, unlike case (i) where it is always less than  $V_l$. This enhancement in the average velocity is because of $\langle k_{\rm on}(x) \rangle$ being larger than $k^0_{\rm on}$ by a factor of by $L/\sqrt{b^2-a^2}$; multiplying $k^0_{\rm on}$ by this factor can collapse both data onto a single curve which stays below $V_l$ (Fig.~\ref{fig:active_np}(c)). Nevertheless this suggests that 
if confinement were to cause an enhancement of the average binding rate $\langle k_{\rm on}(x) \rangle$, it would lead to larger average velocities of the MCC. In fact, in a study similarly to (ii) but for Brownian ratchets in confined media, the authors found an enhancement of the net particle for non-processive motors with a confinement dependent binding rate~\cite{Malgaretti2012,Malgaretti2013}. 
 The qualitative behavior of the effective diffusion coefficient for an MCC with spatially varying binding rates was observed to be similar to that for an MCC with constant binding rates (results not shown here).

\subsection{Reduced Probability Density for Bound and Unbound States}

%\footnote{Because of the channel periodicity and open boundary conditions in our model, $P_{\rm b}(x,t)$ and $P_{\rm ub}(x,t)\to 0$ at long times and do not provide any useful information about the steady states of system.} 

To understand the behavior of the scaled average velocity, in particular the observed data collapsed, we examine the probability densities of the bound and unbound states. 
Given the periodic nature of the corrugated confining channel \cite{footnote_prob}, we study the reduced probability densities \cite{Reimann2002} for the bound state $\hat{P}_{\rm b}(x,t)$ and unbound state $\hat{P}_{\rm b}(x,t)$ 
defined as follows: 
\begin{subequations}
\begin{eqnarray}
\label{eqn:reducedprob}
\hat{P}_{\rm b}(x,t) &=& \sum_{n=-\infty}^{n=\infty} P_{\rm b}(x + nL,t),\\
\hat{P}_{\rm ub}(x,t) &=& \sum_{n=-\infty}^{n=\infty} P_{\rm ub}(x + nL,t), 
\end{eqnarray}
\end{subequations}
with $\int_0^Ldx[\hat{P}_{\rm b}(x,t)+\hat{P}_{\rm ub}(x,t)]=1$. We have checked numerically that these probability densities reach their steady state values $\hat{P}^{st}_{\rm b}(x)$ and $\hat{P}^{st}_{\rm ub}(x)$ at large times.

\begin{figure}[]
\centering
\mbox{
\hspace{-1cm}
\includegraphics[angle=-90,width=0.33\textwidth]{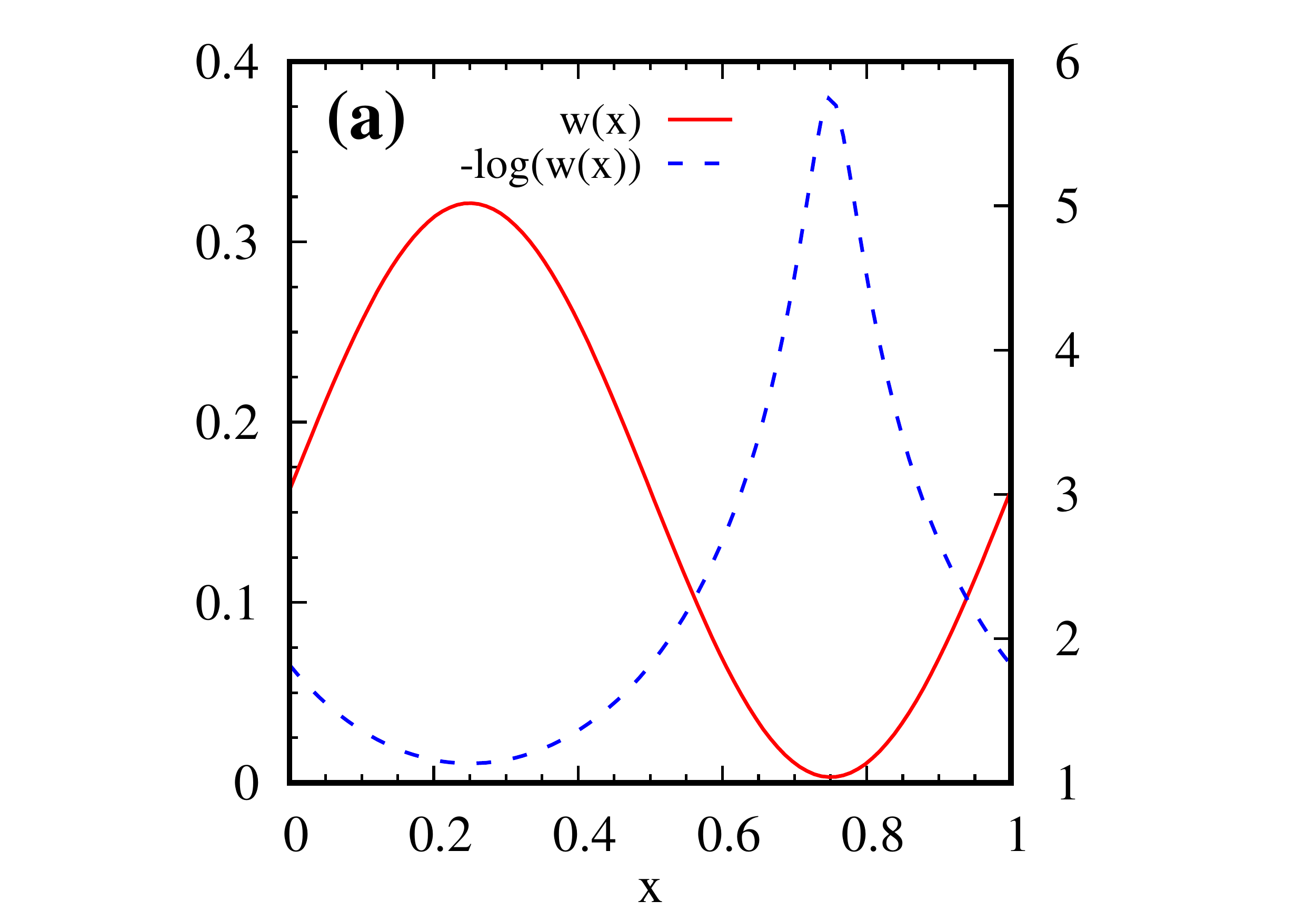}
\hspace{-1.5cm}
\includegraphics[angle=-90,width=0.33\textwidth]{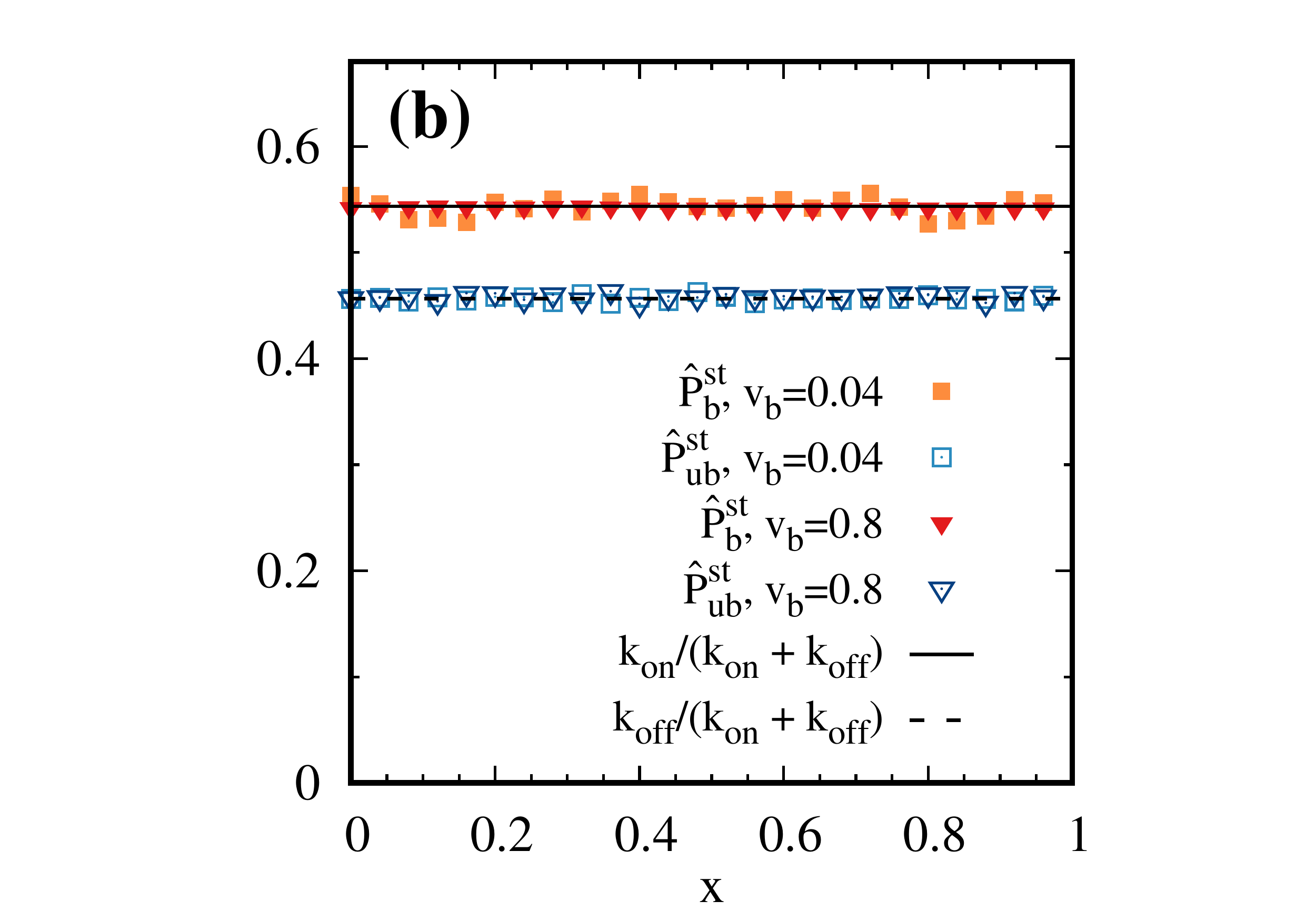}
}
\mbox{
\hspace{-1cm}
\includegraphics[angle=-90,width=0.33\textwidth]{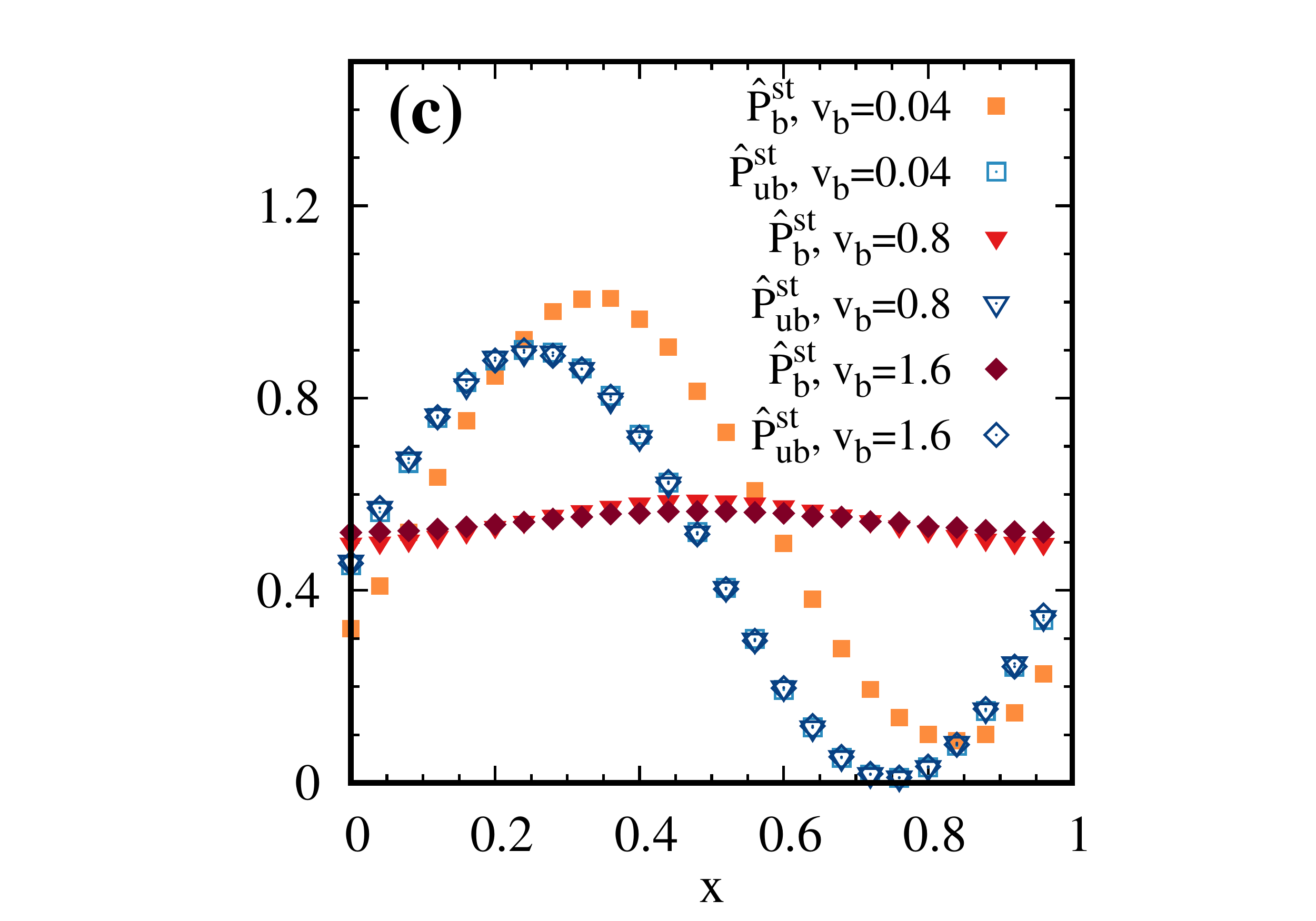}
\hspace{-1.5cm}
\includegraphics[angle=-90,width=0.33\textwidth]{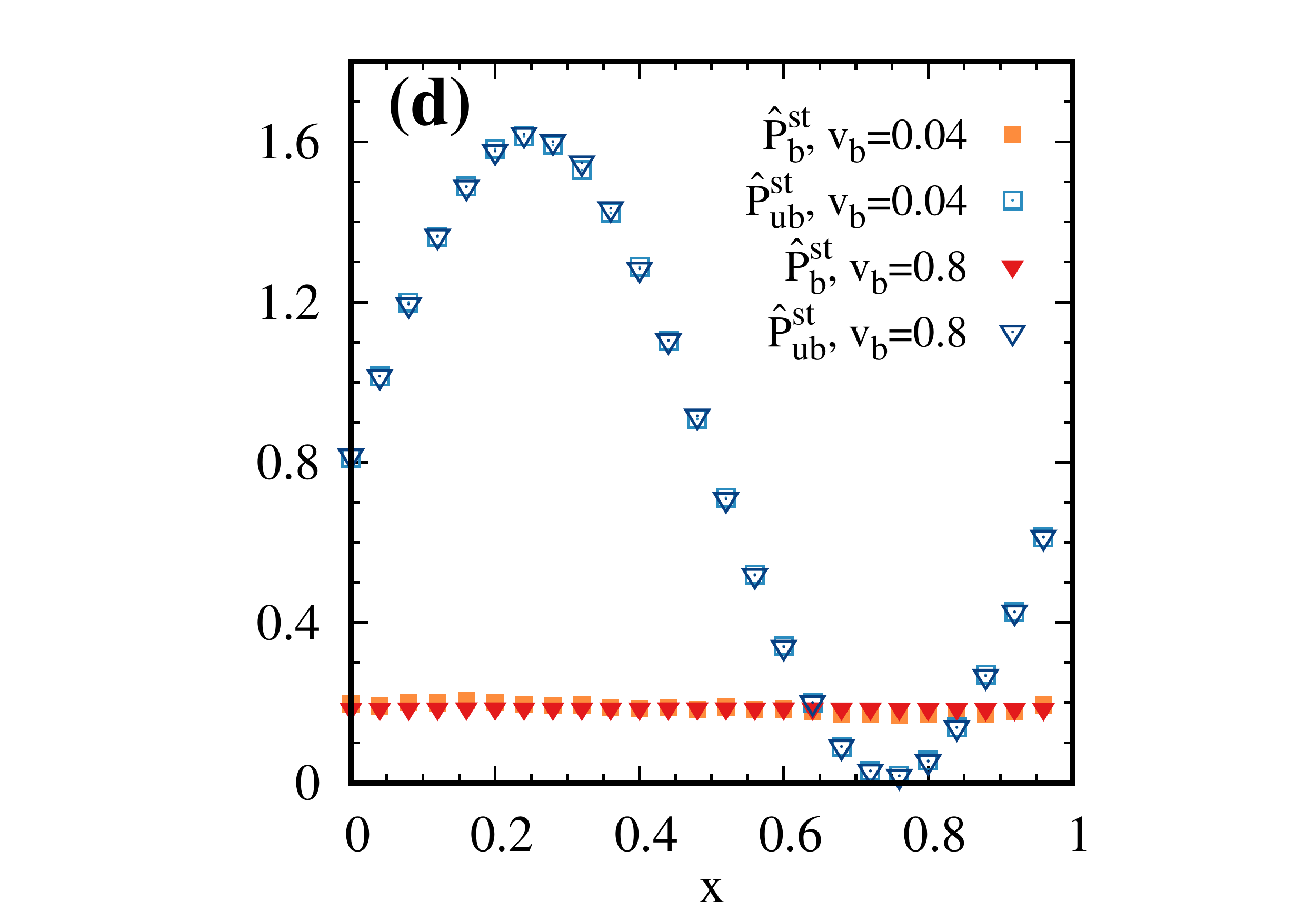}
}
\caption{(a) The wall function $w(x)$ with $L=1$ and the entropic barrier due to $w(x)$.  Figures (b), (c), and (d) show the steady state reduced probability densities $\hat{P}^{st}_{\rm b}(x)$ and $\hat{P}^{st}_{\rm ub}(x)$. Data is shown for small and large $v_b$ ($0.04$, and $0.8 \,\mu m \, s^{-1}$), while $D_0 = 0.64 \,{\mu m^2} s^{-1}$, $k^0_{\rm on}=0.5\,s^{-1}$, and $k_{\rm off}=0.42\,s^{-1}$. Figure (b) shows data for the unconfined case, figure (c) for a confined MCC with constant binding rates, and figure (d) for a confined MCC with that has spatially varying binding rates scaled by the confinement $w(x)$.}
\label{fig:probdist}
\end{figure}

In Fig.~\ref{fig:probdist}, we have shown $\hat{P}^{st}_{\rm b}(x)$ and $\hat{P}^{st}_{\rm ub}(x)$ for three cases: (i) unconfined (Fig.~\ref{fig:probdist}(b)), (ii) confinement with constant binding rate with $k_{\rm on}=k_{\rm on}^0$ (Fig.~\ref{fig:probdist}(c)), and (iii) confinement with spatially varying binding rate with $k_{\rm on}(x) = k_{\rm on}^0\sqrt{(b^2-a^2)}/(L\,w(x))$ (Fig.~\ref{fig:probdist}(d)). In Fig.~\ref{fig:probdist}(a), we plot the spatial profile of the confining wall $w(x)$ and its entropic barrier. For case (i), $\hat{P}^{st}_{\rm b}(x)$ and $\hat{P}^{st}_{\rm ub}(x)$ are uniform in $x$, independent of $v_b$, and follow the corresponding analytical predictions as expected. Interestingly, for case (ii), the $\hat{P}^{st}_{\rm b}(x)$ and $\hat{P}^{st}_{\rm ub}(x)$ are not uniform but modulate with the same wavelength as that associated with the spatial variation of the confining channel. The probability density $\hat{P}^{st}_{\rm ub}(x)$ is independent of $v_b$, while $\hat{P}^{st}_{\rm b}(x)$ vary with $v_b$, approaching to a constant value at large $v_b$. For case (iii), $\hat{P}_{\rm ub}^{st}$ follows the spatial variation of the channel in $x$, while $\hat{P}_{\rm b}^{st}$ is uniform-- this is because the normalization of the binding rate involves scaling by $w(x)$; both these reduced probability densities are independent of $v_b$. It is important to note that for the latter case the value of $\hat{P}_{\rm ub}^{st}$ ($\hat{P}_{\rm b}^{st}$) has significantly increased (decreased) (Fig.~\ref{fig:probdist}(c)) compared to that for the confined MCC with constant binding rate (fig.~\ref{fig:probdist}(b)). This enhancement of $\hat{P}_{\rm ub}^{st}$ and decrease in $\hat{P}_{\rm b}^{st}$ leads to smaller scaled velocity in the case of spatially varying (normalized) binding rate.

In summary, when the reduced probability densities are independent of $v_b$, the corresponding scaled velocity data will collapse onto a single curve, and vice versa. This also explains why for the case of confinement with constant binding rate, scaled velocity data collapse was only observed for large $v_b$, but not for small $v_b$. Furthermore, the increase in unbinding probability density in the case of spatially varying binding rate explains the decrease in the scaled velocity (in Fig.~\ref{fig:active_np}).

\section{Summary and Discussion} 
We have studied the role of confinement in two-state cargo transport in a two-dimensional corrugated channel using the Fick-Jacobs formalism, and an equivalent one-dimensional lattice model. The effect of confinement is incorporated through a position dependent entropic barrier. At any given time, the MCC can be in one of two states: an active state where it moves on a microtubule track with a constant speed, and a passive state when it is detached from the microtubule and undergoes diffusive motion. We assumed small cargo sizes such that while the diffusive motion is impaired by confinement, the bound state directed motion is not. The results from the lattice model exactly match known analytical results for purely diffusive motion in confinement, demonstrating that the Arrhenius description for hopping rates works for our system and other similar systems with entropic barriers. Moreover, the lattice based approach and simple evolution rules make our model computationally more efficient for simulating two state transport in complex confinement profiles than numerical simulations of the corresponding 2D Langevin equations.

In order to understand and quantify how confinement impacts transport properties, we computed and compared the MSD, as well as the average velocity and effective diffusivity of the MCC with and without confinement. The MSD of the confined MCC shows three distinct dynamical regimes corresponding to diffusive motion at small times, ballistic motion at large times, and sub-diffusive motion at intermediate times. The crossover timescale ($\sim 10^{-2}\,s$) from diffusive to sub-diffusive motion is determined by the interplay between passive or diffusive motion and confinement, and suggests a mesh size $\sim 100\,nm$ for the parameters used in our study if the confinement were due to a cytoskeletal network. The crossover from sub-diffusive to the ballistic motion is dictated by the motor properties, specifically the binding kinetics and the speed of the motor when bound. Confinement significantly reduces the crossover time from diffusive to sub-diffusive behavior, and also leads to a significant intermediate sub-diffusive regime. For unconfined MCCs, this intermediate regime is either absent or much smaller than for confined MCCs.

We also found that confinement effectively enhances the motor unbinding rate and thus reduces the average velocity when the bound velocity is small, but has a negligible effect otherwise.  
The impact of confinement on the effective diffusivity is more remarkable. In the absence of any confinement, for less active MCCs ($v_b \lesssim 0.8\,\mu m\,s^{-1}$ ),  an increase in the binding rate leads to a decrease in the effective diffusivity because of the comparatively less time spent in the unbound state; for more active MCCs  ($v_b \gtrsim 1.6\, \mu m\,s^{-1}$), however, the diffusivity initially increases with the binding rate reaching a peak, and then decreases. This can be attributed to locked-to-running transitions in the two state model. While confinement leads to smaller diffusivities, the peaks now start appearing at smaller $v_b \sim 0.8\,\mu m\,s^{-1}$ and are more prominent. Since kinesin-1 motors have an in vitro speed of  $0.8\,\mu m\,s^{-1}$ and an in vivo speed of $2.0 \,\mu m\,s^{-1}$~\cite{bionumbers}, the peaks should be readily observed in experiments in live cells. In vitro, the predictions of our model can be tested in experiments on kinesin-based microtubule transport in enclosed microfluidic channels~\cite{Huang2005}. 

An exception to the above confinement induced slowing down of the MCC is observed when confinement enhanced the average rate of binding of the MCC to the microtubule, thereby leading to an enhancement in the average velocity. 
This suggests that the impact of confinement on cargo transport strongly depends on if and how it modulates the binding kinetics of the motors. Its impact on binding kinetics can be obtained in enclosed microchannel experiments~\cite{Huang2005} by measuring the MCC residence times in the bound and unbound states for different channel widths. 

The same experimental set up can be used to obtain the scaled average velocity, and thus test the predictions of our study. In addition to studying motor driven cargo transport in confinement, such microfabricated enclosed channels can be potentially used to deliver specific proteins or to separate DNA or RNA strands from a complex mixture by binding them to microtubules and transporting them to desired locations. Our results therefore may not only be useful in understanding cargo transport in cells, but also may help in advancing the nanoscale drug delivery system within cells and sequencing techniques for DNA and RNA. Our model can also be easily extended to study bidirectional cargo transport~\cite{bidirectional}. 

Finally, for completeness, we comment on the effect of hydrodynamic coupling between the wall and the cargo in the light of a recent experimental study \cite{Yang2017}.  In our work, we study only the effect of entropic barrier ignoring the hydrodynamic coupling between the confining channel and the diffusing particle, as the latter is not taken into account in the Fick-Jacobs approach \cite{Reguera2001,Reguera2006,Malgaretti2012,Malgaretti2013}.  A recent experiment on colloidal diffusion in corrugated micro-channels found that confinement can increase the hydrodynamic drag which is not captured by the Fick-Jacobs theory using free diffusivities \cite{Yang2017}. However, the authors have demonstrated that this theory can be used to explain their results if it is reformulated in terms of the experimentally measured diffusion coefficients. It may be interesting to study how hydrodynamic effects impact our system; while it is outside the scope of our current study, we will pursue this in future work. Such effects have been found to be important for microswimmers which, unlike diffusive particles, create and use hydrodynamic flow fields for their propulsion \cite{Malgaretti2017}.

\begin{acknowledgments}
The authors would like to thank Jennifer Ross, Megan Valentine, and Ajay Gopinathan for illuminating and helpful discussions, and acknowledge helpful suggestions from anonymous reviewers. SD would also like to thank Dibyendu Das for useful discussions. This research is funded in part by the Gordon and Betty Moore Foundation through Grant GBMF5263.02 to MD. MD and KC were also partially supported by  a Cottrell College Science Award from Research Corporation for Science Advancement.
\end{acknowledgments}

\appendix
\section{Continuous limit of the Lattice model}
\label{app:a}

As discussed in the main text, the master equations describing the time evolution of the probability densities of the bound (active) and unbound (passive) state are given by,
\begin{widetext}
\begin{eqnarray}
\label{eqnapp:master_b}
\frac{\partial P_{\rm b}(x,t)}{\partial t} &=& k_{\rm on}(x) P_{\rm ub} (x, t) - k_{\rm off}(x) P_{\rm b} (x, t) + \lambda_v P_{\rm b} (x - \ell, t) - \lambda_v P_{\rm b} (x, t)\\\nonumber
\label{eqnapp:master_ub}
\frac{\partial P_{\rm ub}(x,t)}{\partial t} &=& -k_{\rm on}(x) P_{\rm ub} (x, t) + k_{\rm off}(x) P_{\rm b} (x, t) + \lambda^{\rm ub}_{+} (x - \ell) P_{\rm ub} (x - \ell, t) + \lambda^{\rm ub}_{-}(x + \ell) P_{\rm ub} (x + \ell, t) \\& & - \left(  \lambda^{\rm ub}_{+}(x) + \lambda^{\rm ub}_{-}(x)\right) P_{\rm ub} (x, t),\\\nonumber
\end{eqnarray}
where $k_{\rm on}(x)$ and $k_{\rm off}(x)$ are the transition rates for the bound and unbound state respectively, $\lambda_v=v_b/\ell$ is the hopping rate in the forward direction when the motor-cargo complex is in the bound state, and $\lambda^{\rm ub}_{\pm}(x) = (D_0/\ell^2){\rm e}^{-\beta(\mathcal{A}(x \pm \ell) -\mathcal{A}(x))/2}$ is the unbound state hopping rates for the forward and backward direction respectively. Using Taylor's expansion for  $P_{\rm b,ub} (x \pm \ell, t)$ and $\lambda^{\rm ub}_{+} (x \pm \ell)$ around $x$ and keeping the terms up to $2^{nd}$ order in $\ell$, we get 
\begin{eqnarray}
\label{eqnapp:approx_b}
\frac{\partial P_{\rm b}(x,t)}{\partial t} &=& k_{\rm on} P_{\rm ub} - k_{\rm off} P_{\rm b} - v_b \frac{\partial P_{\rm b}}{\partial x} + D_v \frac{\partial^2 P_{\rm b}}{\partial x^2} + \mathcal{O}(\ell^3),\\ \nonumber
\label{eqnapp:approx_ub}
\frac{\partial P_{\rm ub}(x,t)}{\partial t} &=& -k_{\rm on} P_{\rm ub} + k_{\rm off} P_{\rm b}
+ \ell \, P_{\rm ub} \left[ - \frac{d\lambda^{\rm ub}_{+}}{dx} + \frac{d\lambda^{\rm ub}_{-}}{dx} + \frac{\ell}{2} \left(\frac{d\lambda^{\rm ub}_{+}}{dx} - \frac{d\lambda^{\rm ub}_{-}}{dx} \right) \right] \\ 
& & + \ell\, \frac{\partial P_{\rm ub}}{\partial x} \left[ -\lambda^{\rm ub}_{+} + \lambda^{\rm ub}_{-} + \ell ( \frac{d\lambda^{\rm ub}_{+}}{dx} + \frac{d\lambda^{\rm ub}_{-}}{dx} ) \right] + \frac{\ell^2}{2} \frac{\partial^2 P_{\rm ub}}{\partial x^2}  \left( \lambda^{\rm ub}_{+} + \lambda^{\rm ub}_{-} \right) + \mathcal{O}(\ell^3),
\end{eqnarray}
where $D_v\equiv\ell^2 \lambda_v/2=\ell\,v_b/2$. In $\ell \rightarrow 0$ limit, neglecting the term with $D_v$ in Eq.~\ref{eqnapp:approx_b} we recover the continuum Fokker-Planck equation for the bound state (Eq.~(2a) in the main text)
\begin{eqnarray}
\label{eqnapp:approx_b}
\frac{\partial P_{\rm b}(x,t)}{\partial t} &=& k_{\rm on} P_{\rm ub} - k_{\rm off} P_{\rm b} - v_b \frac{\partial P_{\rm b}}{\partial x}.\\ \nonumber
\label{eqnapp:approx_ub}
\end{eqnarray}
Considering the leading order contributions for the coefficients of $P_{\rm ub}$ and $\frac{dP_{\rm ub}}{dx}$ in Eq.~\ref{eqnapp:approx_ub} we get
\begin{eqnarray}
\label{eqnapp:approx1_ub}
\frac{\partial P_{\rm ub}(x,t)}{\partial t} &=& -k_{\rm on} P_{\rm ub} + k_{\rm off} P_{\rm b}
+ \ell \, P_{\rm ub} \left[ - \frac{d\lambda^{\rm ub}_{+}}{dx} + \frac{d\lambda^{\rm ub}_{-}}{dx} \right]  
 + \ell\, \frac{\partial P_{\rm ub}}{\partial x} \left[ -\lambda^{\rm ub}_{+} + \lambda^{\rm ub}_{-} \right] + \frac{\ell^2}{2} \frac{\partial^2 P_{\rm ub}}{\partial x^2}  \left( \lambda^{\rm ub}_{+} + \lambda^{\rm ub}_{-} \right).
\end{eqnarray}
For $\ell \rightarrow 0$ limit, $\lambda^{\rm ub}_{\pm}(x) = (D_0/\ell^2) {\rm e}^{\mp \frac{\beta \ell}{2} \frac{d\mathcal{A}}{dx}}$.  In this limit, the coefficients of $P_{\rm ub}$, $\frac{\partial P_{\rm ub}}{\partial x}$, and $\frac{\partial^2 P_{\rm ub}}{\partial x^2}$ are given by,  $\ell \left( -\frac{d\lambda^{\rm ub}_{+}}{dx} + \frac{d\lambda^{\rm ub}_{-}}{dx} \right) \simeq D_0 \beta \frac{d^2\mathcal{A}}{dx^2}$,
$\ell\left(-\lambda^{\rm ub}_{+} + \lambda^{\rm ub}_{-} \right) \simeq  D_0 \beta \frac{d\mathcal{A}}{dx}$, and $\ell^2 \left(\lambda^{\rm ub}_{+} + \lambda^{\rm ub}_{-}\right)/2 \simeq D_0$ respectively. Using these expressions in Eq.~\ref{eqnapp:approx1_ub} we recover the continuum Fokker-Planck equation for the unbound state (Eq.~(2b) in the main text),
\begin{eqnarray}
\nonumber
\frac{\partial P_{\rm ub}(x,t)}{\partial t} &=& -k_{\rm on} P_{\rm ub} + k_{\rm off} P_{\rm b}  + D_0 \beta \frac{\partial }{\partial x} \left(P_{\rm ub} \frac{d\mathcal{A}}{dx}\right) + D_0 \frac{\partial^2 P_{\rm ub}}{\partial x^2}\\
&=& -k_{\rm on} P_{\rm ub} + k_{\rm off} P_{\rm b} + D_0 \frac{\partial }{\partial x}\left(e^{-\beta \mathcal{A}(x)} \frac{\partial}{\partial x} e^{\beta \mathcal{A}(x)}P_{\rm ub}(x,t)\right).
\label{qnapp:approx2_ub}
 \end{eqnarray}

\end{widetext}

%

%\bibliography{cargo_transport1}

\begin{thebibliography}{59}
\bibitem [{foo()}]{footnote_prob}%
  \BibitemOpen
  \href@noop {} {}\bibinfo {note} {Because of the channel periodicity and open
  boundary conditions in our model, $P_{\rm b}(x,t)$ and $P_{\rm ub}(x,t)\to 0$
  at long times and do not provide any useful information about the steady
  states of system.}\BibitemShut {Stop}%
\end{thebibliography}
%\bibliographystyle{h-physrev}

\end{document}